%% file: ms.tex
\documentclass[conference]{IEEEtran}
\IEEEoverridecommandlockouts

\usepackage[utf8]{inputenc}
\usepackage{amsmath}
\usepackage{color}
\usepackage{graphicx}
\usepackage{url}
\usepackage{siunitx}
\usepackage{listings}
\usepackage{multirow}
\usepackage{enumitem}
\usepackage{subfig}
\usepackage{graphicx}
\usepackage{pgfplots}

\usepackage[acronym]{glossaries}
\setacronymstyle{long-short}
\newacronym{st}{S/T}{Schmitt-Trigger}

\usepackage{textpos}
\usepackage{hyperref}
\usepackage[capitalize]{cleveref}

\input{common.tex}

\begin{document}

\title{Efficient Metastability Characterization for Schmitt-Triggers}
\author{
  \IEEEauthorblockN{ J\"urgen Maier
    \begin{minipage}[c]{1em}
      \href{https://orcid.org/0000-0002-0965-5746}{{\includegraphics[width=1em]{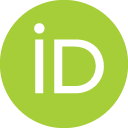}}}
    \end{minipage}\,
    and\: Andreas Steininger
    \begin{minipage}[c]{1em}
      \href{https://orcid.org/0000-0002-3847-1647}{{\includegraphics[width=1em]{orcID.png}}}
    \end{minipage}
  }
\IEEEauthorblockA{TU Wien, 1040 Vienna, Austria\\
\{jmaier, steininger\}@ecs.tuwien.ac.at}
\thanks{This research was partially supported by the SIC project (grant
  \mbox{P26436-N30}) of the Austrian Science Fund (FWF).}
}

\maketitle

\begin{textblock*}{\textwidth}(0mm,189mm)%
  \footnotesize%
  \textcopyright\ 2019 IEEE.  Personal use of this material is permitted.  Permission from IEEE
  must be obtained for all other uses, in any current or future media, including
  reprinting/republishing this material for advertising or promotional purposes,
  creating new collective works, for resale or redistribution to servers or lists,
  or reuse of any copyrighted component of this work in other works.
\end{textblock*}%
\vspace*{-1em}

\begin{abstract}
  Despite their attractiveness as metastability filters, Schmitt-Triggers can
  suffer from metastability themselves. Therefore, in the selection or
  construction of a suitable Schmitt-Trigger implementation, it is a necessity
  to accurately determine the metastable behavior.  Only then one is able to
  compare different designs and thus guide proper optimizations, and only then
  one can assess the potential for residual metastable upsets.  However, while
  the state of the art provides a lot of research and practical characterization
  approaches for flip-flops, comparatively little is known about Schmitt-Trigger
  characterization. Unlike the flip-flop with its single metastable point, the
  Schmitt-Trigger exhibits a whole range of metastable points depending on the
  input voltage. Thus the task of characterization gets much more challenging.

  In this paper we present different approaches to determine the metastable
  behavior of Schmitt-Triggers using novel methods and mechanisms. We compare
  their accuracy and runtime by applying them to three common circuit
  implementations.  The achieved results are then used to reason about the
  metastable behavior of the chosen designs which turns out to be problematic in
  some cases.  Overall the approaches proposed in this paper are generic and can
  be extended beyond the Schmitt-Trigger, i.e., to efficiently characterize
  metastable states in other circuits as well.
\end{abstract}

\section{Introduction}
\label{sec:introduction}

\begin{figure}[t]
  \centering
  \includegraphics[width=0.9\linewidth]{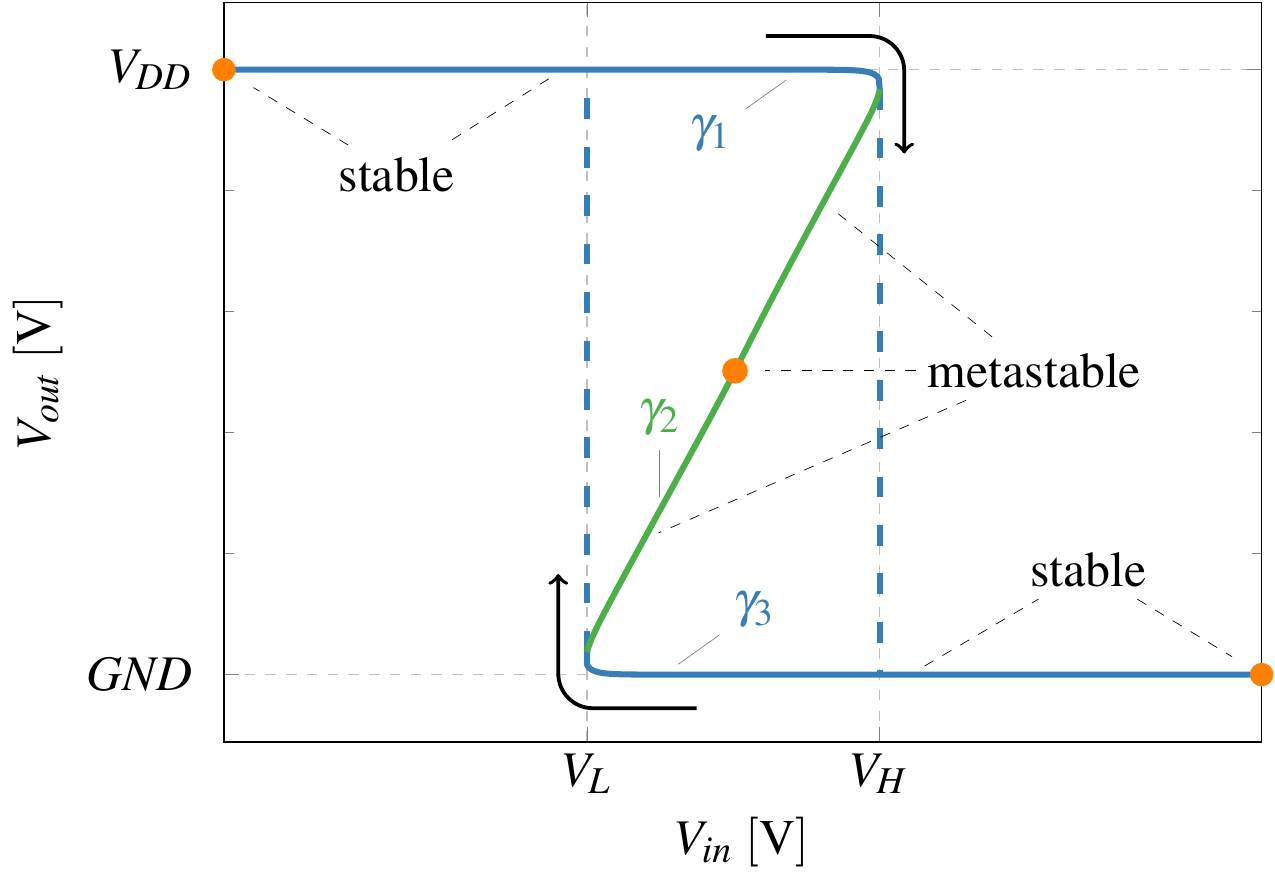}
  \caption{Stable and metastable states of a latch (dots) and an  \gls{st}
    (lines) forming the characteristic z-shaped curve.}
  \label{fig:meta_states}
  \vspace{-0.3cm}
\end{figure}

To use digital abstraction in electronic circuits we have to ``digitize'' an
essentially analog input, i.e. either assign logic HI or LO depending on whether
it is above or below a threshold.  In order to prevent oscillation of the output
due to noise in case of an input voltage close to the threshold -- like in case
of a comparator circuit --, the \gls{st} uses a higher threshold for rising
transitions than for falling ones, leading to a hysteresis (blue lines in
Fig.~\ref{fig:meta_states}).  This, however, directly translates into a
dependence of the threshold on the current output state, which, in turn, implies
a positive feedback from the output to the input. As a consequence, the \gls{st}
must be susceptible to metastability.  This intuitive argument has been more
formally supported by Marino~\cite{Marino77} already, and more recently
Steininger et al.~\cite{SMN16:ASYNC} have detailed several practically relevant
scenarios where metastability may occur and where it may not. While there exist
analytic solutions to calculate certain properties such as the threshold
voltages~\cite{Dokic12}, none have been presented so far regarding
metastability.

Actually \gls{st} metastability is detrimental to its popular use for
``cleaning'' noisy input signals, or conditioning metastable outputs produced by
other elements. Therefore it is crucial to fully characterize the metastable
behavior of an \gls{st} and, in the ideal case, estimate a mean time between
(metastable) upsets (MTBU), as it is common with metastability in
flip-flops. The latter has been well researched since the seminal work by
Kinniment et al.~\cite{KE72}, Chaney et al.~\cite{CM73}, and
Veendrick~\cite{Vee80}. However, as it turns out, the \gls{st} case
substantially differs by the fact that its input remains connected to the
positive feedback loop all the time, which ultimately results in the \gls{st}
exhibiting a whole range of metastable voltages $\vmeta(\vin)$ rather than just
one as in the flip-flop case. This makes the analysis and characterization way
more complicated, and, unfortunately, hardly any results are available, apart
from the mentioned papers~\cite{Marino77} and~\cite{SMN16:ASYNC}.

Before it is possible to analyze the traces and probabilities to enter and leave
these metastable states, and thus achieve a similar expression like the MTBU of
a flip-flop, we need an efficient and accurate approach to characterize a given
implementation. In~\cite{SMN16:ASYNC} the authors used transient analog
simulations to search for the metastable values, a very time consuming procedure
which includes lots of manual steps yet. In addition they solely showed their
analysis on a single circuit, a six-transistor implementation for a \SI{65}{\nm}
technology. Therefore it is not clear whether the results are transferable to
other implementations and technologies as well.

\textbf{Contribution: } While we cannot solve all the open problems mentioned
above in this paper, we extend the work by Steininger et
al. in~\cite{SMN16:ASYNC} by presenting and critically analyzing different
approaches to simulate/evaluate the metastable behavior of an \gls{st}. This not
only includes the (meta-)stable states but also the behavior in their
surrounding. These data are of interest when investigating more advanced
features such as the overall probability to enter metastability or how quickly
it is resolved. More specifically, we

\begin{itemize}
\item derive a more fine grained map (compared to~\cite{SMN16:ASYNC}) of
  the output derivative $\vout'$ over the $\vin-\vout$ plane, which
  we use as basis for more accurate estimations and analyses about the general
  behavior.
\item determine all stable states, which partly lie in the undefined voltage 
  range\footnote{We call the voltage range above a well-defined LO and below a
    well-defined HI, according to the logic specification, undefined. In a
    properly functioning (metastability-free) circuit this range is crossed by
    steep transitions only.} for certain implementations
  making metastable behavior very easily reachable.
\item carry out a preliminary analysis on metastability resolution, which turns out be
  shaped exponentially (comparable to the flip-flop) but with varying parameter
  $\tau(\vin)$.
\item introduce a novel method, which is not limited to \glspl{st}, that makes
  stable points metastable and vice versa.
\item exploit plain \dc\ analyses to determine metastability.
\item evaluate the single approaches by characterizing three common
  implementations and comparing the results among each other and with the
  analytic results from~\cite{Marino77}.
\end{itemize}

This paper is organized as follows: In section~\ref{sec:background} we briefly
review metastability of \glspl{st} followed by a description of the proposed
characterization methods in section~\ref{sec:metastability}. Results and
discussion for three common implementations are shown in
section~\ref{sec:circuits} which is followed by a conclusion and an outlook to
future research possibilities in section~\ref{sec:conclusion}.

\begin{figure}[t]
  \centering
  \includegraphics[width=0.75\linewidth]{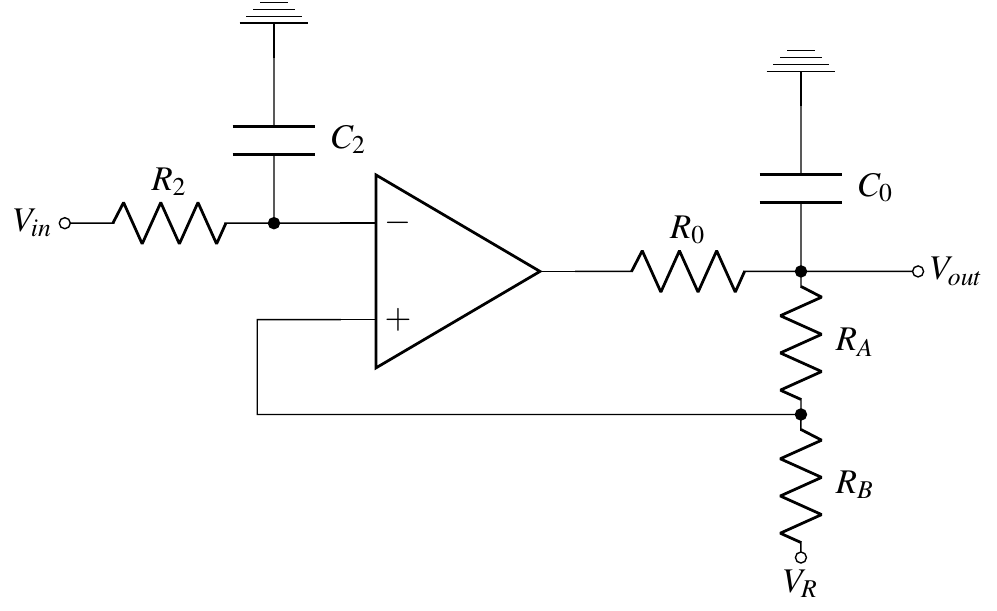}
  \caption{Schmitt Trigger implementation studied in~\cite{Marino77}.}
  \label{fig:dyn_model}
\end{figure}

\section{Background}
\label{sec:background}

Metastability has been well researched for latches, formed by cross-coupled
inverters, since the seminal work by Kinniment et al.~\cite{KE72} and
Veendrick~\cite{Vee80}. For these elements there is even an equation for the
mean time between metastable upsets (MTBU) available which relies, besides
others, on the metastable resolution constant $\tau_C$.  The amount of stable
(two) and metastable (one) states is very small which is a direct consequence of
the decoupled input. Fig.~\ref{fig:meta_states} shows the single states (dots),
where $\vin$ represents the value inside the loop.

\subsection{Schmitt-Trigger Metastability}
\label{sec:STmeta}

\begin{figure}[t]
  \centering
  \includegraphics[width=0.75\linewidth]{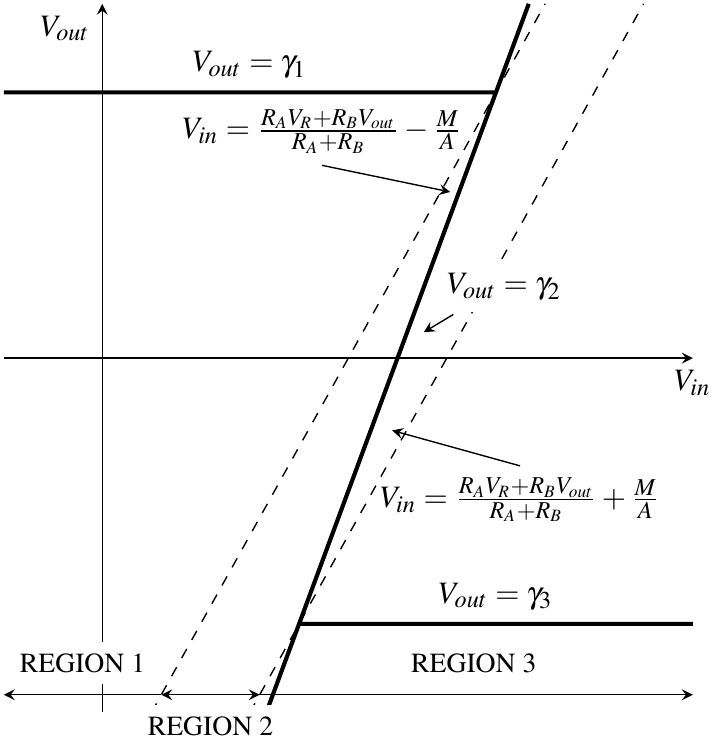}
  \caption{Phase diagram of the \gls{st} inspired by Marino \cite{Marino77}}
  \label{fig:phase_diag}
\end{figure}

Unfortunately the situation is much more complicated for the \gls{st} since the
input remains connected continuously and hence has to be considered as well.
For this purpose Marino~\cite{Marino77} modeled the \gls{st} by a properly wired
OpAmp (shown in Fig.~\ref{fig:dyn_model}) and carried out analytic
considerations. He used the phase diagram ($\vout'$ over the $\vin$--$\vout$
plane, as shown in Fig.~\ref{fig:phase_diag} with $A$ equal to the amplifier
gain and $M$ being the output saturation voltage) to divide the behavior in
three different regions, where the output in each is governed by the following
equations:

\begin{align}
\text{Region~1:}&&
\frac{dV_{out}}{dt} &= V_{out}' = -\frac{1}{\tau_1} (V_{out} - \gamma_1) \\
\text{Region~2:}&&
\frac{dV_{out}}{dt} &= V_{out}' = \frac{1}{\tau_2} (V_{out} - \gamma_2) \\
\text{Region~3:}&&
\frac{dV_{out}}{dt} &= V_{out}' = -\frac{1}{\tau_3} (V_{out} - \gamma_3)
\end{align}

The functions $\gamma_1$ and $\gamma_3$ represent the stable states while
$\gamma_2$, which connects the former, the metastable ones. In contrast to the
latch there are now infinitely many (meta-)stable values ranging from the lower
(\gnd) continuously to the upper ($\vdd$) supply voltage. As was shown
in~\cite{SMN16:ASYNC} with the proper input signal (exceeding the threshold and
then steering back) any of these values can be reached and held forever.

\subsection{Schmitt-Trigger Metastability Characterization}
\label{sec:characterization}

The phase diagram as proposed by Marino is like a finger print of a
Schmitt-Trigger implementation and helps the designer to understand and optimize
the circuit.  Therefore, in this paper we are searching for ways to determine
the phase diagram in a fast, simple and yet accurate fashion for
state-of-the-art implementations. As analytic considerations are based on
certain abstractions, which are a good way to recognize dependencies however
lack accuracy especially for modern technologies, we decided to base our
analysis on \spice\ simulations.

\subsubsection{Stable States (\hyst)}
\label{sec:hysteresis}

Let us first focus on $\gone$ and $\gthree$. These are very easy to achieve by
starting two \dc\ analyses, one sweeping $\vin$ from \gnd\ to $\vdd$ and one in
the opposite direction. The threshold voltages, $\vl$ and $\vh$, are easily
recognized, as a small change on $\vin$ leads to a major jump on $\vout$. Please
note that, in contrast to the analysis of Marino, $\gone$ and $\gthree$ are
neither constant functions, nor straight lines in real circuits. Instead the
stable values start to deviate from (\gnd/$\vdd$) when the threshold voltage is
approached (cp. Fig.~\ref{fig:meta_states}). For certain implementations this
change is substantial, as we will show in Section~\ref{sec:circuits}, and thus
has to be carefully analyzed. This is even more important as these states are
actually stable and thus much easier to reach than metastable ones on $\gtwo$,
i.e., simply by a rising input stopping at a specific value
(cp. \cite{SMN16:ASYNC}).

\subsubsection{Metastable States}
\label{sec:characteristics}

Far more interesting for us is however $\gtwo$. Simply connecting $\gone$ and
$\gthree$ by a straight line, as derived by Marino, yields a first
approximation, for more accurate result we have to resort however to more
advanced methods. Luckily metastable states can be uniquely identified by
checking for $\vout'=0$, a property that all points on $\gone$, $\gtwo$ and
$\gthree$ share. This immediately follows from the fact that one can stay
infinitely long in perfect metastability and of course stable states.

Steiniger et al.~\cite{SMN16:ASYNC} used transient analysis for this purpose. In
detail they observed for a pair of $\hat{\vin}$ and $\hat{\vout}$ if $\vout(t)$
in- or decreased during a simulation run. Based on the result they implemented a
binary search algorithm for the value of $\hat{\vout}$ in the next simulation
until the desired accuracy was achieved. This procedure was then repeated for
numerous values of $\vin$ along $\gtwo$.  Since this is a very time consuming
task we searched for more ingenious solutions and even found several
alternatives, which we will describe in the following.

\section{Methods for obtaining $\gtwo$}
\label{sec:metastability}

Efficient and precise ways to determine the (metastable) characteristics of a
\glsdesc{st} are key for reliability predictions or comparisons between
different implementations. In the following we will elaborate several approaches
to determine the metastable states ($\gtwo$), as this is currently the biggest
challenge.  For accurate simulations we resorted to \hspice\ using a
\SI{28}{\nm} UMC technology library. Comparisons with an older \SI{65}{\nm}
technology showed no qualitative difference so we restrict ourselves to
presenting the former in this paper. Our circuit model is pre-layout, but we
consider a capacitive output load of $C_L=\SI{2}{\fF}$ in our \ac\ analyses.

\subsection{Static Analysis of Grid Points (\map)}
\label{sec:dVout}

Recall from Section~\ref{sec:characteristics} that all (meta-)stable states
share the property $\vout'=0$. As a first approach we can cover the
$\vin$-$\vout$ plane with a regular grid and determine $\vout'$ for each grid
point to find where it gets (close to) zero.  Albeit this initially appears
quite untargeted and laborious, it provides us with a map that will turn out
valuable for analyzing the resolution behavior later on.

\begin{lstlisting}[caption=deriving $\iout$ in $\vin$-$\vout$ plane in \spice, label=lis:dVout]
.DC VIN 0 supp width SWEEP VOUT LIN count 0 supp
.PROBE DC I(VOUT)
\end{lstlisting}

\begin{figure}[t]
  \centering
  \includegraphics[width=0.9\linewidth]{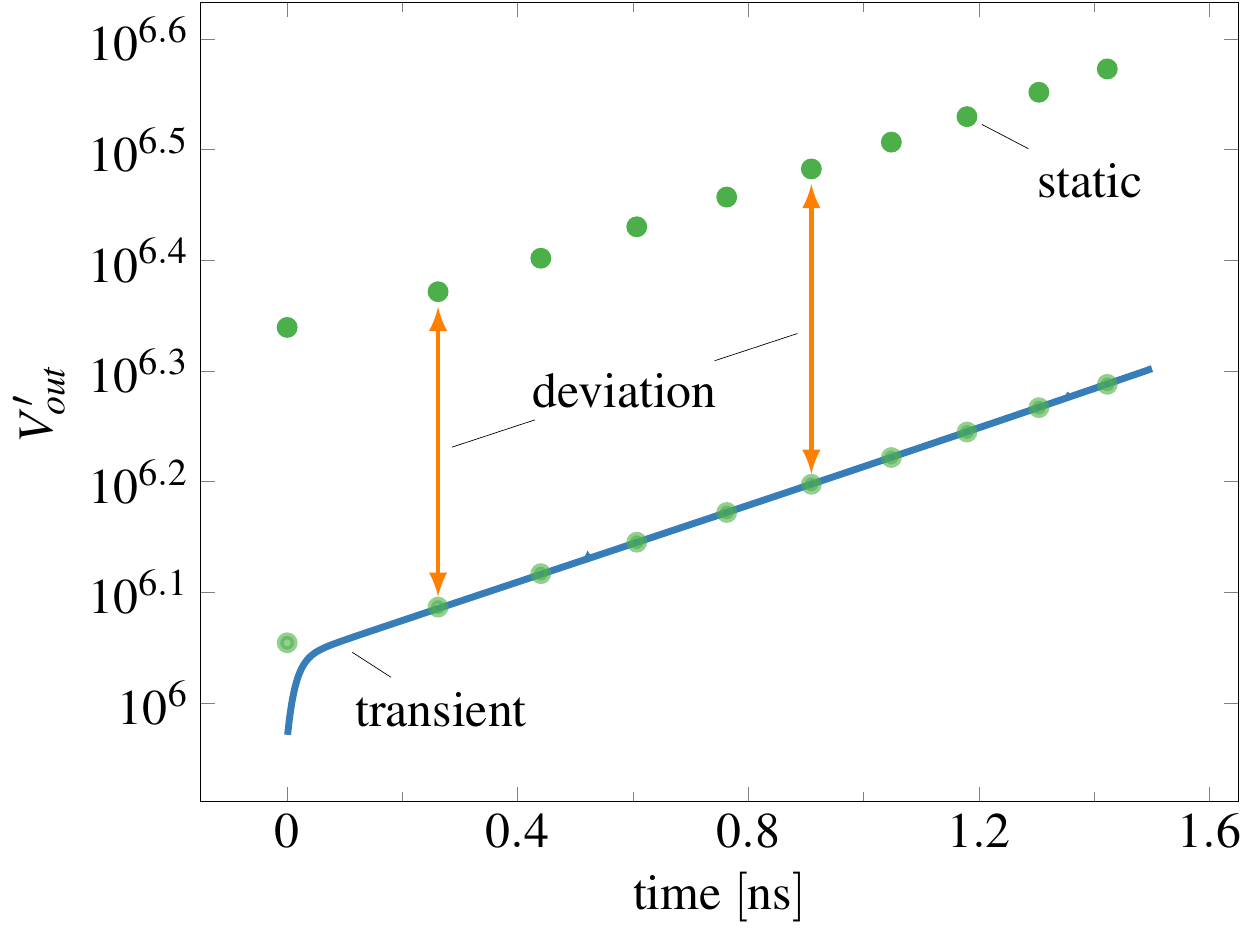}
  \caption{Output derivative for transient and static simulations. The deviation
    stays constant as the projection (opaque green dots) shows.}
  \label{fig:transient_sim}
\end{figure}

Our approach uses built-in commands from \spice\ only, as detailed in
Listing~\ref{lis:dVout}: We sweep $\vin$ from $0$ to $\vdd$ (\emph{supp}) in
steps (\emph{width}) corresponding to the grid. In the same way $\vout$ is swept
(\emph{count} $=$ \#steps). For this purpose we replace the load capacitance by
a voltage source and actually measure the current through the latter (visible in
the second code line).  In comparison to the approach from~\cite{SMN16:ASYNC},
where the authors performed transient analysis and picked $\vout'$, this is
considerably faster but serves the same purpose, albeit we get $\iout$ as a
result instead.  To compare the results of static and transient simulation (see
Section~\ref{sec:transient}) in Fig.~\ref{fig:transient_sim} we can use the
transformation $\vout' \cdot C_{L} = \iout$, which leads however to a constant
deviation. That discrepancy is a result of the internal capacitance of the
\gls{st} which we determined to be \SI{1.854}{\fF} and whose value stays
constant even for varying values of $C_L$. In the following we will therefore
use $\hat{C_L}=C_L+\SI{1.854}{\fF}$ for transformations between $\vout'$ and
$\iout$.

Note that our proposed \dc\ analysis does not reflect potential dynamic effects:
In real circuits $\vin'$ most certainly has an effect on $\vout'$ through
coupling capacitances. However, in our view this only restricts the possible
paths a metastable state can be reached, but not the actual value
itself, since all (meta-)stable states are per definition static, i.e., $\vout'=0$.
Therefore we consider it fundamental to determine the
static, general case in the first place -- and this is what the \dc\ analysis properly does.

Obviously we won't be lucky enough to hit $\iout=0$ (or $\vout'=0$) exactly this
way, but those pairs of grid points between which $\iout$ changes its sign
already confine $\gtwo$.  In a first step contour plots can be used to draw an
(interpolated) contour line at $\iout=0$, i.e., at the (meta-)stable
line. Furthermore the map may serve as starting point for more precise
estimations.

\subsection{Transient Estimation (\tranEst)}
\label{sec:transient}

When starting transient simulations in the grid points confining $\gtwo$
(changing sign of $\iout$) one observes traces that are nearly perfect
exponentials (see Fig.~\ref{fig:transient_sim}), as predicted by
theory~\cite{Marino77}.  This has major implications. Firstly, it suggests that
the resolution behavior is comparable to the flip-flop, with the main
difference, however, that the resolution time constant $\tau$ is now not unique
but varies with $\vin$.  This will become apparent in
Section~\ref{sec:circuits}.

Secondly, it gives us the possibility to infer the metastable voltage by
recording just a short piece.  Assume we start in an arbitrary point
$(\hat{V}_{in},\hat{V}_{out})$ and observe the output, i.e., $\vout(t)$ and
$\vout'(t)$ in the course of the simulation.  Since we assume the trace to be
exponential we get the following relations:

\begin{align}
  \label{eq:Vout}
 \vout = \vmeta \pm V_x \cdot \exp \left( \frac{t-\hat{t}}{\tau}
 \right)  \\
   \label{eq:dVout}
 \vout' = \pm  \frac{1}{\tau}\cdot V_x \cdot \exp \left( \frac{t-\hat{t}}{\tau} \right)
\end{align}

\noindent where $\hat{t}$ denotes the unknown time shift between our measurement
and the actual resolution curve and $V_x>0$ the unknown scaling factor of the
exponential.  Let us now apply the natural logarithm on $|\vout'|$ leading to

\[ \ln (|\vout'|) = \ln \left[\frac{1}{\tau} \cdot V_x \cdot \exp \left(
      -\frac{\hat{t}}{\tau} \right) \right] + \frac{t}{\tau} \]

By applying a linear fit to our simulation results we can easily determine 
the value of $\tau$, which is inversely proportional to the slope.
Going back to Equation~(\ref{eq:Vout}) and 
expressing the exponential term by $\vout'$ finally yields:

\[ \vmeta = \vout \mp \tau \cdot \vout' \]

$\vmeta$ is obtained by plugging in a single pair of measured values for $\vout$
and $\vout'$. Note that we actually do not have to know the absolute time (or
the parameters $\hat{t}$ and $V_x$, respectively): We used the time and value
difference between some measured values of $\vout'$ to determine $\tau$, while a
consistent pair of $(\vout, \vout')$ sufficed to finally obtain $\vmeta$.

For valid results, two aspects have to be considered:
\begin{itemize}
\item Initially, $\vout'$ observed by the transient analysis changes
  disproportionately (cp. Fig.~\ref{fig:transient_sim}) leading to a bad
  fitting. As a consequence the first samples have to be removed.

\item At some point the waveform changes from ``leaving metastability''
  (increasing $|\vout'|$) to ``approaching stable value'' (decreasing $|\vout'|$). By
  choosing initial conditions and simulation time it must assured that this point is
  never reached by the simulation.
\end{itemize} 

\subsection{Static Estimation (\dcEst)}
\label{sec:static}

In accordance to the transient measurements of $\vout'$ we also see an
exponential growth of the static $\iout$ as we follow a resolution trajectory,
which is reasonable as they only differ by a constant factor.  Therefore it is
quite natural to apply the same estimations as before also on $\iout$.  One can
rewrite the expressions in Equation~(\ref{eq:dVout}) to

\begin{align}
  \label{eq:Iout}
\vout' = \frac{\iout}{\hat{C_L}} =  \frac{1}{\tau} \cdot  (\vout - \vmeta)
\end{align}

$\hat{C_L}/\tau$ can be achieved by the slope of $\iout$ over $\vout$, i.e., by
fitting the data from the $\iout$ map.
%
Plugging an arbitrary $\vout$, the corresponding $\iout$ and the known
$\hat{C_L}/\tau$ into Equation~(\ref{eq:Iout}) finally yields the metastable
voltage $\vmeta$.

Compared to the transient analysis the calculations on $\iout$ are far easier to
execute and thus less prone to errors. Both provide however the possibility
to improve the limited accuracy of tools (due to numerical issues).

\begin{figure}[t]
  \centering
  \includegraphics[width=0.9\linewidth]{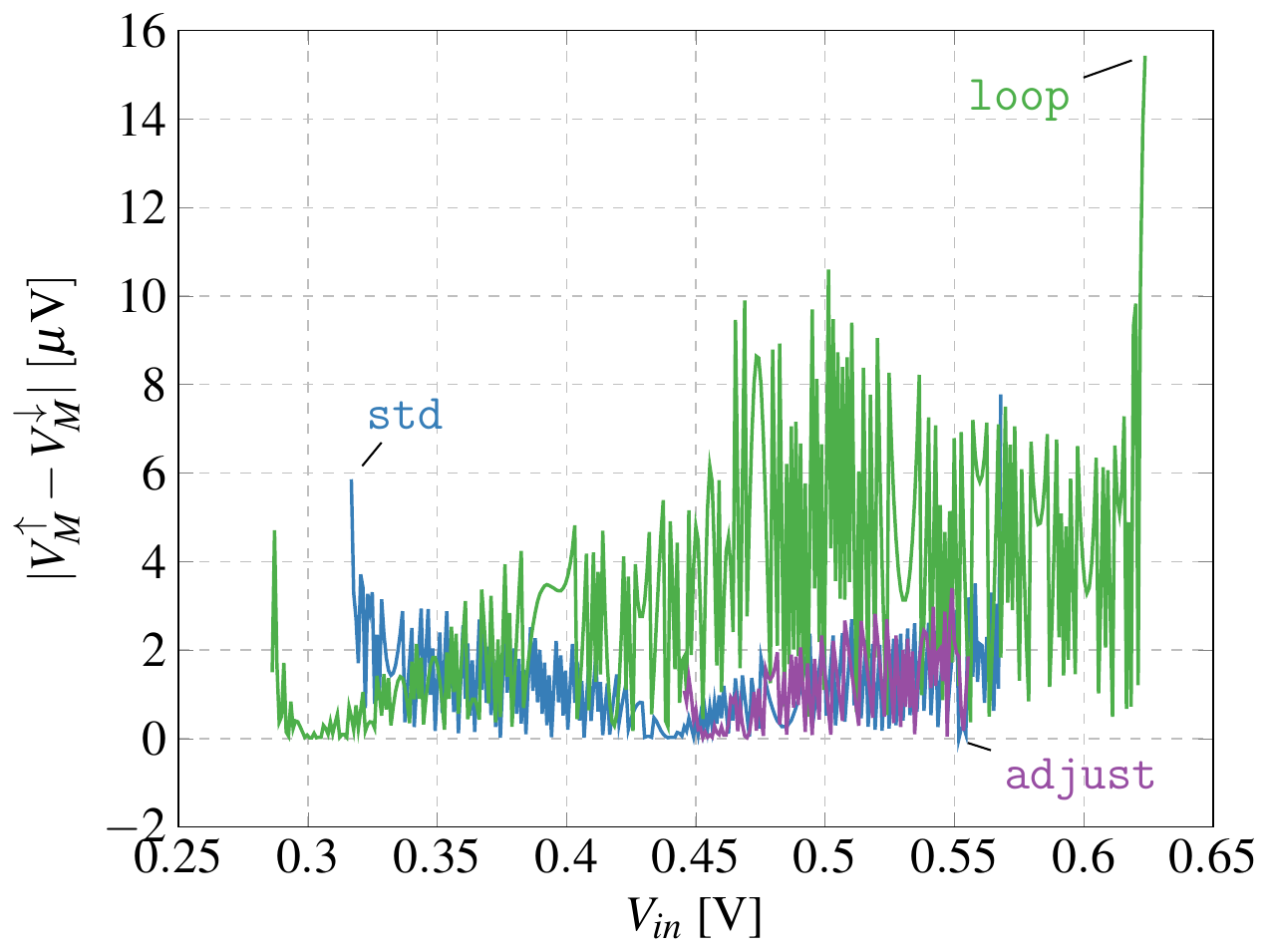}
  \caption{Absolute deviation between $\vmeta$ predictions based on the resolution
    direction for different circuit implementations (cp. Section~\ref{sec:circuits}).}
  \label{fig:dev_vmeta}
\end{figure}

Please note that for both, \tranEst\ and \dcEst, $\tau$ and $\vmeta$ can be
determined twice: either for traces resolving the metastability to \gnd\, or for
such resolving to $\vdd$. Ideally both would render the same results.  In
reality, however, we get slightly different values for $\tau$ (which may indeed
have a physical reason), and slight deviations in $\vmeta$ (most likely due to
numerical issues, which have an exponential effect). For the latter
Fig.~\ref{fig:dev_vmeta} shows the difference between the predictions for the
curve resolving to $\vdd$ ($\vmeta^\uparrow$) and $\gnd$ ($\vmeta^\downarrow$)
for \dcEst, whereat results for \tranEst\ are slightly worse. For better
accuracy we therefore determine $\vmeta$ (using the corresponding $\tau$) as the
point where the two resolution curves meet.

\subsection{Binary Search (\binary)}
\label{sec:binary}

A more pragmatic approach is \binary, where we sweep $\vin$ from $\vl$ to $\vh$,
and for each value a binary search is performed to find a fitting value of
$\vout$, i.e., where $\iout$ is zero. This is very similar to the approaches
in~\cite{SMN16:ASYNC,YG07:DATE} with the difference that we use the static
current instead of the transient output derivative.

To our advantage \spice\ has a mechanism to run a binary search built-in (called
``Bisection'') which simplifies the task a lot.  The corresponding code is shown
in Listing~\ref{lis:bisection}. The first line states that we want to bisect,
and at most $40$ steps shall be carried out. Note that this narrows down the
initial interval by a factor of $2^{40}$, so most of the time the algorithm
quits earlier, as the demanded accuracy is reached first.

\begin{lstlisting}[caption=bisection in \spice, label=lis:bisection]
.model optMod1 OPT METHOD=BISECTION ITROPT=40
.param outVal=optFunc1(vdd/2, vout_VL, vout_VH)
.DC VIN inVal inVal 1 SWEEP OPTIMIZE=optFunc1
+ RESULTS=optMeasure MODEL=optMod1
\end{lstlisting}

The second line sets the parameter \emph{outVal} which determines the output
voltage and the range within which it shall be swept. We used here the value of
$\vout$ at $\vin=\{\vl,\vh\}$ as we have to be sure to avoid the stable
states. The search itself always starts at $\vdd/2$. The last line finally
launches the \dc\ analysis for the input voltage in the range
[\emph{inVal},\emph{inVal}], meaning that we execute this search for each value
of $\vin$ separately, since we were not able to convince \spice\ to do that
automatically.

\subsection{Metastability Inversion (\invTran)}
\label{sec:meta_inversion}

The reason why metastable states are hard to characterize is the fact that it is
close to impossible to actually reach them even in simulations, since, by
definition, the system consistently works towards leaving them.  A good physical
analogy is the inverted pendulum.  Stable states, in contrast, are naturally
assumed by the system. This observation suggests reduced characterization effort
if the cases could be reverted, i.e., stable points are made metastable and vice
versa.

With this in mind, let us model the system in the metastable state as an
(output) current source that is controlled by the output voltage, namely

\[\iout = K \cdot (\vout-\vmeta)\]

A positive current charges the load capacitance, i.e., increases $\vout$,
 which in turn increases the
current even more. For the voltage gradient we get

\[\vout' = \frac{\iout}{\hat{C_L}} =  \frac{K}{\hat{C_L}} \cdot (\vout-\vmeta)\]

\noindent which yields an exponential function for $\vout$, more specifically,
with $K > 0$ an exponentially growing one.  We can, however, invert the sign
of $K$ by connecting a current source $I_L$ to the output, whose current is
controlled by $\iout$, i.e.

\[I_L = p \cdot \iout.\]

Mathematically, this changes the current into the capacitance from $\iout$ to
$(1-p)  \cdot \iout$, and we get

\[ \vout' = \frac{(1-p) \cdot \iout}{\hat{C_L}} =  \frac{(1-p) \cdot  K}{\hat{C_L}} \cdot (\vout-\vmeta)\]

\noindent which, for $p>1$, yields a decaying exponential function and hence a stable solution.

\begin{figure}[t]
  \centering
  \includegraphics[width=0.7\linewidth]{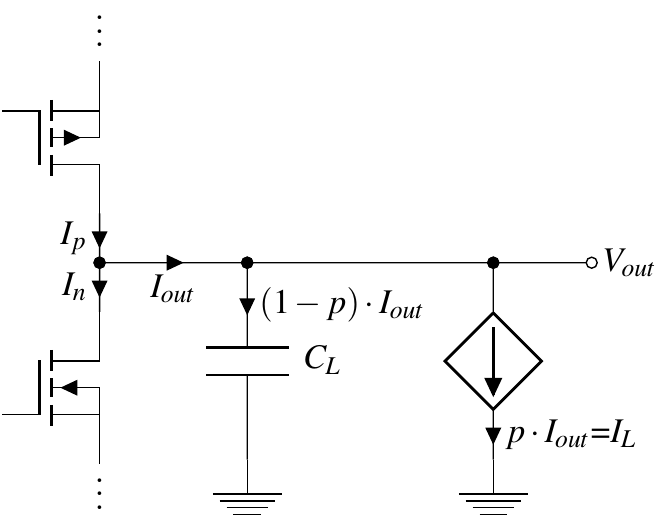}
  \caption{Circuit setup to invert (meta-)stable points.}
  \label{fig:inverted_meta}
\end{figure}

Intuitively spoken, any positive $\iout$ is overcompensated by $I_L$, such that
$C_L$, instead of being loaded by $\iout$, now gets discharged through that
portion of $I_L$ that exceeds $\iout$. Therefore ultimately $\vout$ is reduced,
and, as a consequence, $\iout$ as well.  This naturally drives $\iout$ towards
zero, which represents exactly the state that is normally
metastable. Fig.~\ref{fig:inverted_meta} shows the resulting circuit for
$\iout>0$.

Overall the added current source serves as a proportional controller (for the
current) that stabilizes the ``inverted pendulum''.  Implementing such an
element in \spice, and similarly in other simulation suites, is
straightforward. The only challenge left is to choose a proper value for $p$,
which is a very delicate task: In essence, $p$ contributes to the gain in the
control loop and is hence, according to control theory, crucial for
stability. Choosing $p$ too low (very close to 1) leads to slow stabilization
and consequently long simulation times. Choosing $p$ too high, in contrast,
causes oscillating behavior.

In order to come up with a guideline for a reasonable choice of $p$ we have used
the example of the implementation shown in Fig.~\ref{fig:inverter_loop} that
employs the same inverter loop as the latch. Consequently we could build on the
model from Veendrick~\cite{Vee80}. By extending the latter with our controlled
current source we can approximate the behavior of the dynamic system comprising
\gls{st} and controller. From that it turns out that the right choice for $p$ is

\begin{align} \label{equ:p}
p=2 A^2 \left( 1 + \sqrt{1-\frac{1}{A^2}} \right) \approx 1+\frac{1}{4A^2}
\end{align}

\noindent with $A^2$ being the product of the (\dc) gains of each of the two
inverters in the model.  Note, however, that the inverter gain is not constant,
so the ``ideal'' $p$ will change as we move along $\gtwo$. In our experiments we
chose the highest gain, i.e. the one in the middle of the inverter's transfer
curve, which yielded very useful values for $p$. Luckily \spice\ supports the
determination of the loop gain, so $A^2$ can be obtained by the code piece
shown in \cref{lis:amp}, whereat we connected a constant voltage source to the
input and set the initial value of $\vout$ appropriately. For the complete code refer
to the tool described in \cref{sec:circuits}.

To get as close as possible to the \dc\ case we measure the gain at very low
frequencies (\SI{1e-5}{\Hz}), which is shown in line $3$ in the listing.
Please note that different \gls{st} implementations will lead to different
feedback models, such that \cref{equ:p} cannot be applied directly in these cases.

\begin{lstlisting}[caption=measuring loop gain in \spice, label=lis:amp]
.ac dec '10' '0' '10'
.lstb mode=single vsource=vlstb
.measure lstb gain loop_gain_at_minifreq
\end{lstlisting}

The method of metastability inversion is not restricted to \gls{st} but can
easily be extended to also determine the metastable point of other circuits. We
verified this by introducing a current source into a ``latch-style'' inverter
loop (i.e. with unconnected input) which then quickly approached its metastable
value.

\subsection{DC Analysis (\invDC)}
\label{sec:dc}

\begin{figure}[t]
  \centering
  \includegraphics[width=0.98\linewidth]{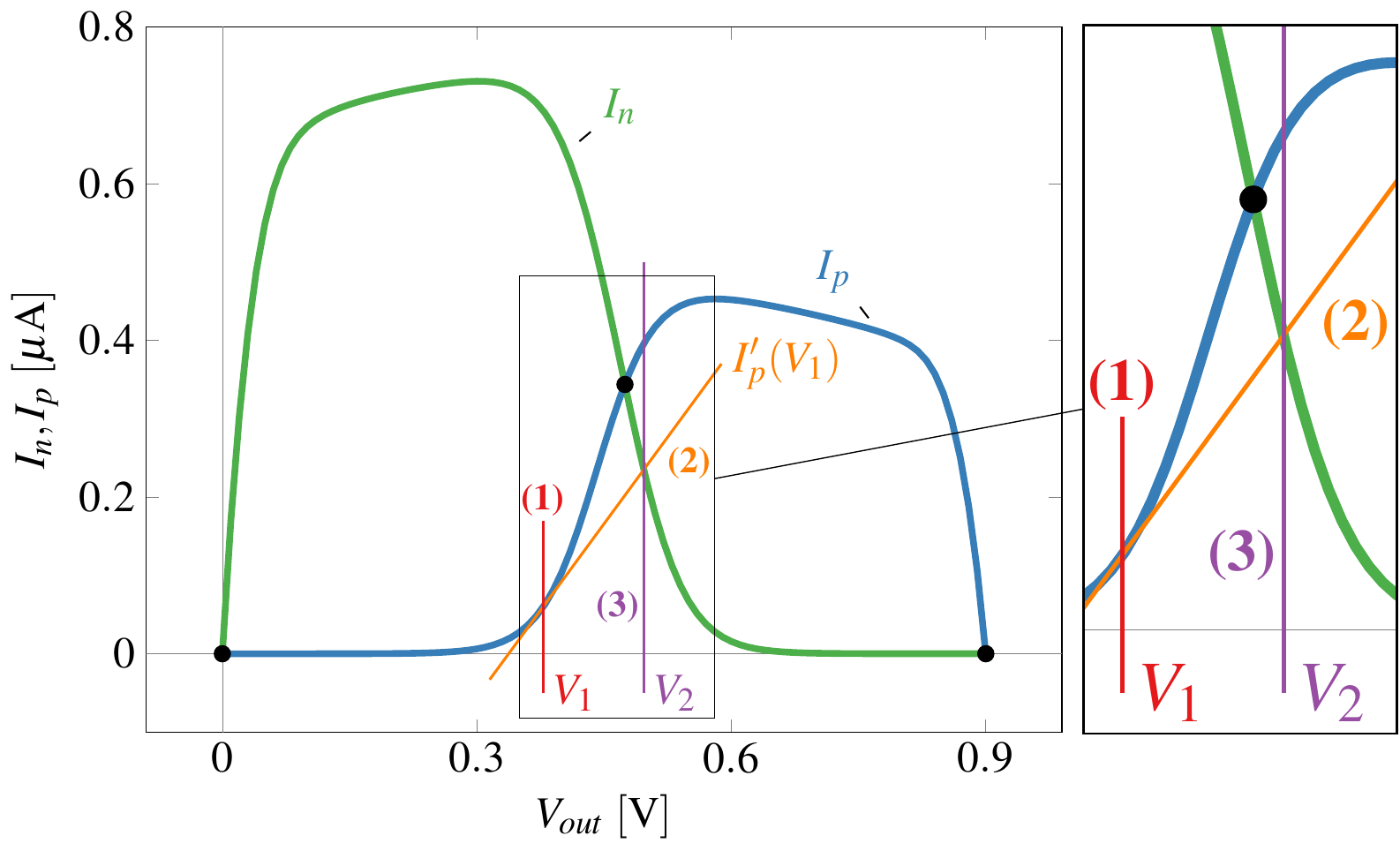}
  \caption{Application of the \emph{Newton-Raphson} algorithm to find a stable
    value (marked by black dot) of $\vout$ for fixed $\vin$.}
  \label{fig:newton}
\end{figure}

With metastability inversion transforming metastable states into stable ones we
could use \dc\ analysis to collect the values on $\gtwo$.  While this works
well, we discovered in the course of our research, that already \dc\ simulations
on the unmodified implementation are capable of delivering the metastable
values. The reason for that is the \emph{Newton-Raphson} algorithm which \spice\
uses to determine the \dc\ operating point \cite{HSPICE_SA}. Let us take a
closer look at this procedure: Assume that we fixed $\vin$ in the metastable
region and want to determine a stable $\vout$ (we already know that there are
exactly three possibilities). For a stable value the current coming from the
p-stack ($\ipmos$) and the one flowing into the n-stack ($\inmos$) at the output
have to be equal. Determining those currents for various values of $\vout$ gives
us traces like those shown in Fig.~\ref{fig:newton}. The three stable states
marked by black dots are clearly visible. To start the algorithm we have to make
a guess, say we pick $V_1$. The next task would be (1) to determine the
derivative of $\ipmos$ in this point, i.e, $\ipmos'(V_1)$, (2) find the crossing
point of the latter with $\inmos$ and finally (3) restart the procedure with
$\vout=V_2$, i.e., the value in the crossing point. If we start close enough to
$\vmeta$ the algorithm will approach it automatically. It can be seen that a
deviation of several tens of millivolts can be tolerated for the initial guess,
an accuracy easily achievable by connecting the last stable values of $\gone$
and $\gthree$ by a straight line.

After the first metastable value was found others follow quickly as
the current value of $\vmeta$ serves as starting point for the search on the
next one, whose value does not differ much and is thus found very rapidly. 

For that reason the most important task is to get a good initial guess for the
first value. If we are too far off, we will find the stable state and thus end
up with part of the hysteresis that we already know.  Assume we start in the
point on $\gtwo$ that is closest to $\gthree$.  Due to this close proximity we
can infer that for that choice the metastable value will be close to the peak
value for $\vout$ on $\gthree$.  Therefore, if we choose a slightly higher value
for $\vout$ we end up with a very good initial guess, since our starting point
is closer to $\gtwo$ than to $\gthree$.  The amount of increase is uncritical
and can actually be chosen rather big, in our experiments up to $\vdd/4$.

\begin{figure}[t]
  \centering
  \includegraphics[width=0.9\linewidth]{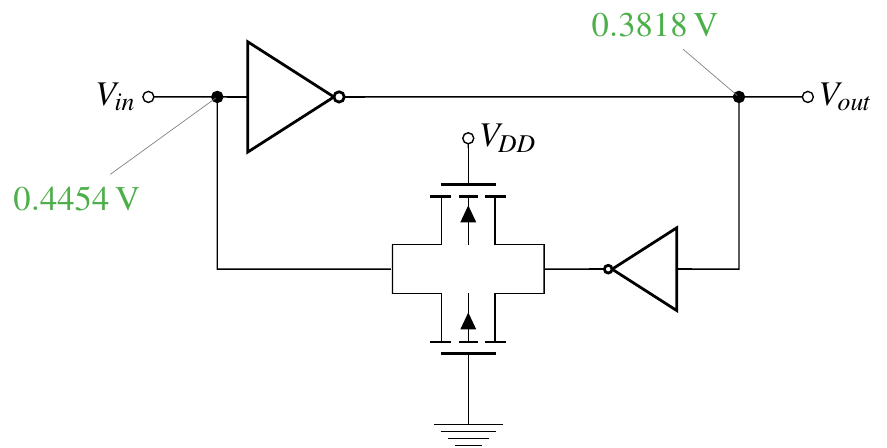}
  \caption{Flip-flop half used for \dc\ metastability analysis.}
  \label{fig:half_ff}
\end{figure}

We varified our approach also on a flip-flop half, i.e., a loop of asymmetric
inverters (width ratio $1/10$) with a transmission gate (see \cref{fig:half_ff})
in a similar fashion. In detail the input values to the first and second
inverter were set to $\vdd/2=\SI{0.45}{\V}$ and then \spice\ was asked to
calculate the operation point. As a result we got $\vin=\SI{0.4454}{\V}$ and
$\vout=\SI{0.3818}{\V}$.

\compPlot{sim_standard6T}{\std}{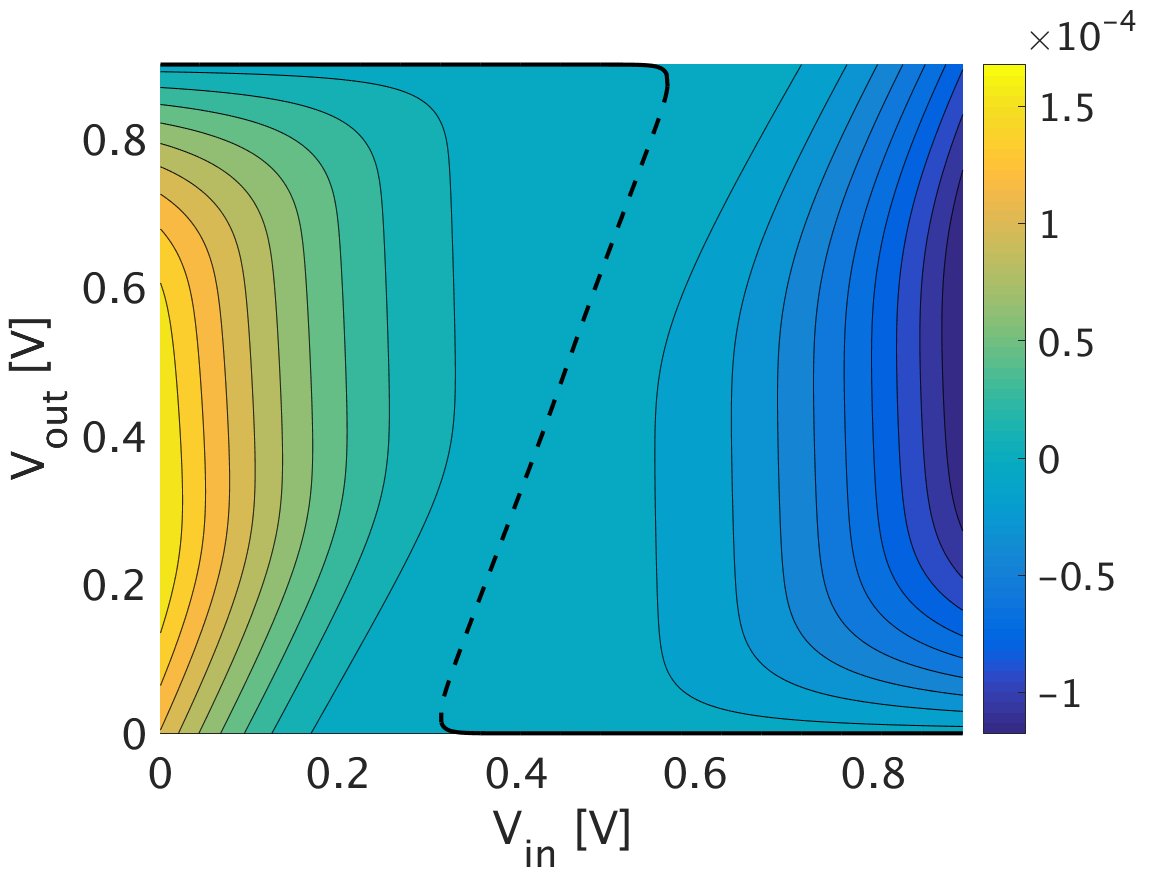}{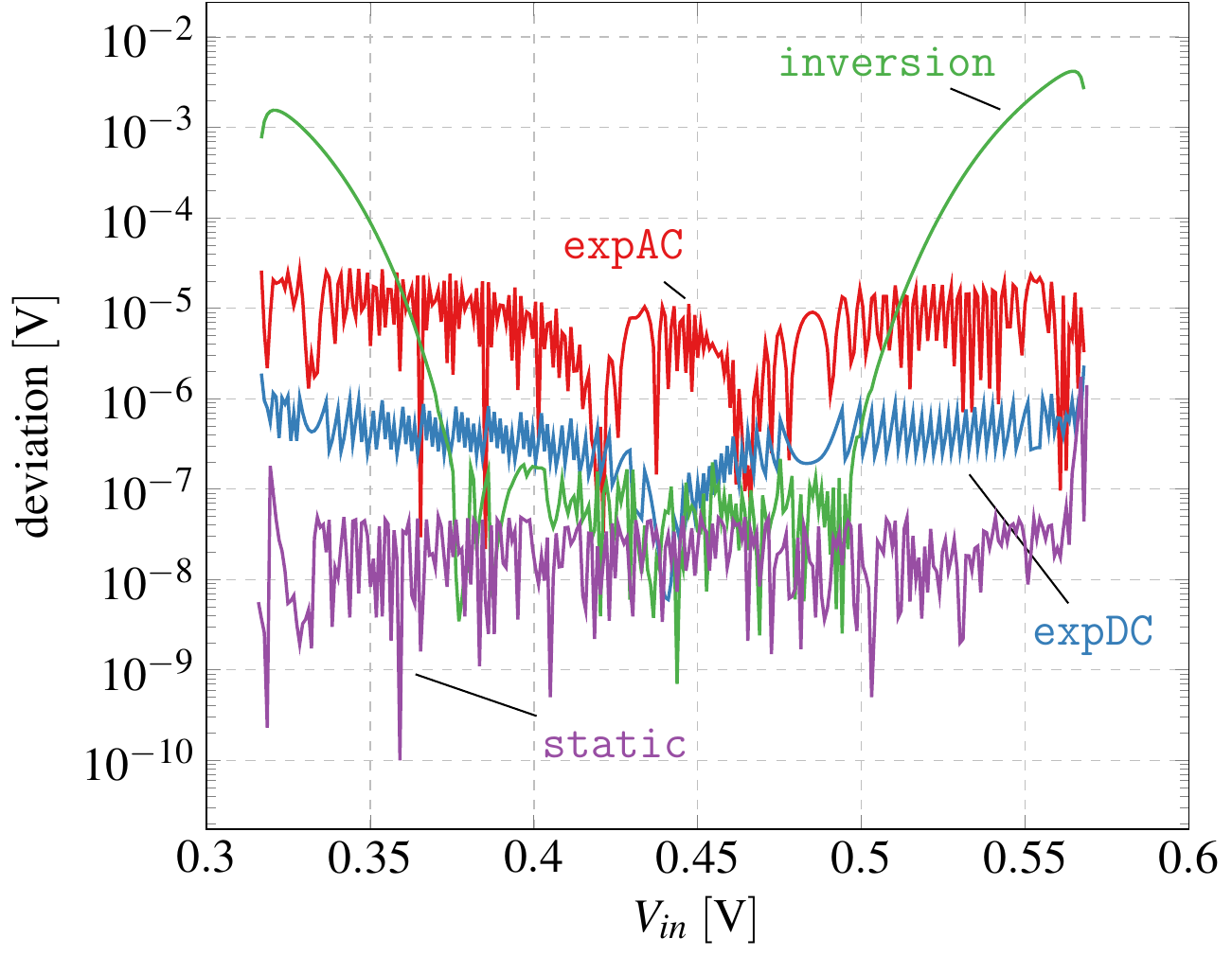}{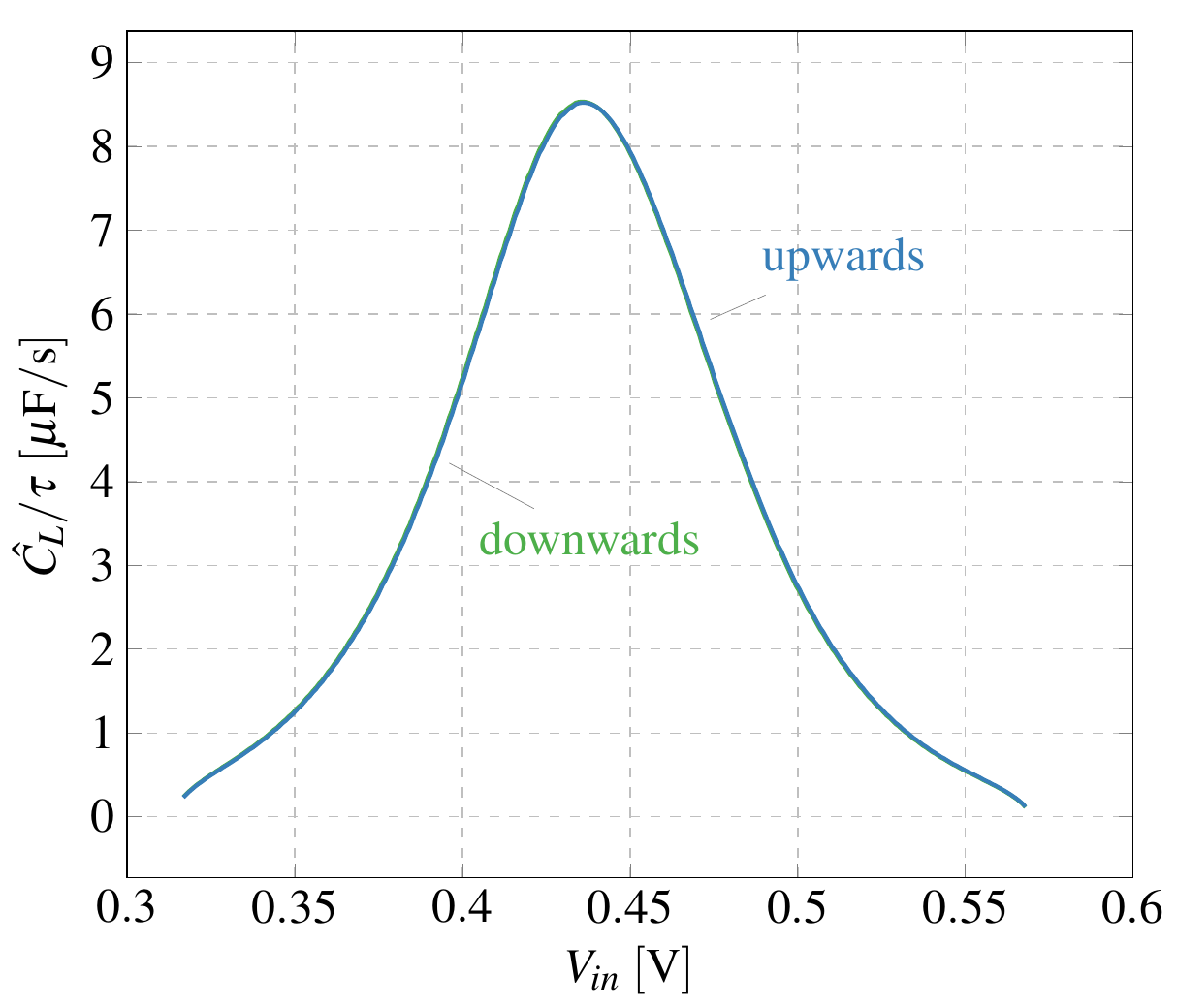}{0.31}

\begin{figure}[b]
  \centering
  \includegraphics[width=0.9\linewidth]{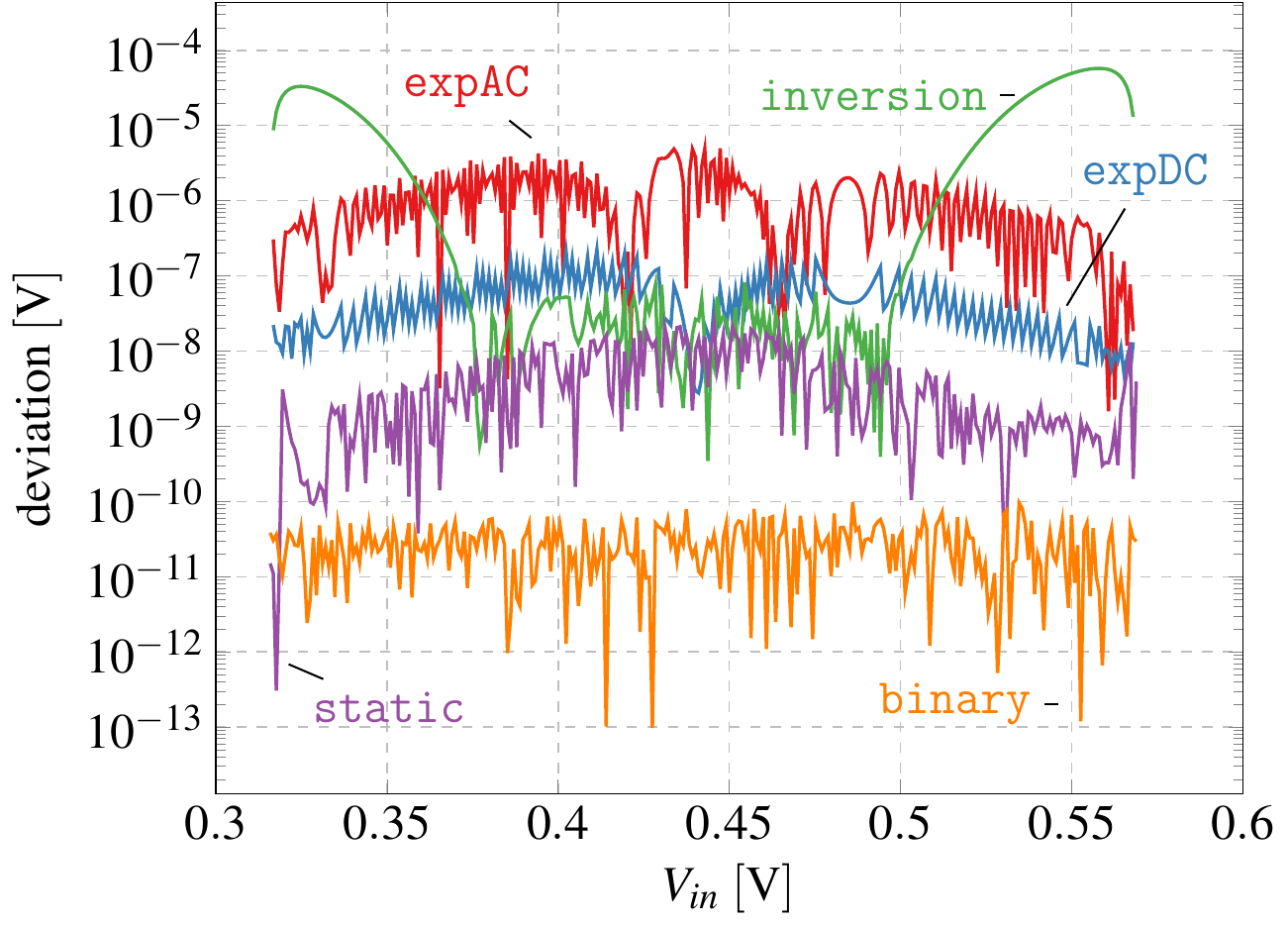}
  \caption{Deviation of $\vout$ after \SI{200}{\ps}.}
  \label{fig:accuracy}
\end{figure}

\section{Evaluation}
\label{sec:circuits}

Provided that an appropriate \spice\ description of the circuit is available,
the complete characterization process can be carried out fully autonomous. Thus
we implemented all the approaches presented in the previous section in a small
tool which is publicly
available\footnote{\textit{https://github.com/jmaier0/meat}}.  For the
simulations presented in the following, we used a \SI{28}{\nm} UMC technology
library ($\vdd=\SI{0.9}{\V}$) and determined the (meta-)stable states for $1000$
equally spaced values of $\vin$.

The aim of these simulations is twofold:
\begin{itemize}
\item We want to evaluate and compare the presented methods in a practical
  application.  To this end we apply them for characterizing three different
  \glspl{st}: a) the standard 6T implementation (\std) b) an inverter
  loop~\cite{BUNDALO1985} (\invLoop) and c) an adjustable hysteresis
  one~\cite{AdjHyst} (\ctrlHyst), as other circuits in literature are heavily
  based on these.

\item The analysis and comparison of the \gls{st} implementations is as such
  important.  In particular it is interesting to see how far the behaviors
  differ among each other and also from the theoretical results~\cite{Marino77}.

\end{itemize}

\subsection{General Remarks}
\label{sec:remarks}

In principle, the \emph{resolution} of all presented methods can be made as high
as desired.  In practice there exist however limitations such as the finite
simulator precision (number format), the required run time and the available
output file formats. The latter caused heavy problems as we only managed to
export results with $7$ positions after the decimal point (solely for \binary\
we achieved $10$ by using a different method).

The \emph{accuracy} of the results is somewhat limited by the achieved
resolution and the assumption of a perfectly exponential resolution trajectory
for \dcEst\ and \tranEst, the simulation time for \invTran, and in general by
the accuracy of the circuit and transistor models underlying the SPICE
simulations (which we will neglect from now on, as this issue is immanent to all
simulation approaches).  To verify that the metastable states obtained through
the various methods are indeed accurate, we started transient simulations in
each of them and measured the output deviation after
\SI{200}{\ps}. Fig.~\ref{fig:accuracy} shows the results for \std.  As one can
see \binary, which predicts $\vmeta$ with a resolution of $\pm \SI{5}{nV}$, has
clearly the lowest deviation and thus is the most accurate. Therefore we will
use \binary\ as golden reference for all further analyses, rather than
continuing with the computationally intensive transient simulations for accuracy
validation.

Where the single methods differ, however, is the \emph{effort} one needs to
invest.  To quantify that, we try to approach the metastable value reasonably
close ($\pm \SI{10}{\uV}$) and then compare the required run times.
Table~\ref{tab:sim_times} gives an overview of the achieved results.  As the
hysteresis differs among the implementations, the number of grid points between
$\vl$ and $\vh$ (metastable grid points), and thus the run time,
varies. Nevertheless, for the same circuit a comparison among different methods
is still valid.

\begin{table}[h!]
  \centering
  \caption{Overview of simulation times}
  \label{tab:sim_times}
  \begin{tabular}{l c S S S}
     & &\multicolumn{3}{c}{simulation time [\si{\s}]} \\
    method && {\std} & {\invLoop} & {\ctrlHyst} \\ 
    metastable grid points && 282& 378& 125\\ \hline \hline
    \hyst && 1.825& 2.012& 1.842\\
    \binary && 232.279& 349.224& 104.585\\
    \map && 310.826& 318.153& 362.692\\
    \tranEst && 526.149& 806.603& 233.369\\
    \dcEst && 2.047& 2.481& 1.310\\
    \invTran && 901.588& 1552.542& 447.089\\
    \invDC && 0.850& 0.939& 0.853\\
  \end{tabular}
\end{table}

\compPlot{sim_inv_loop}{\invLoop}{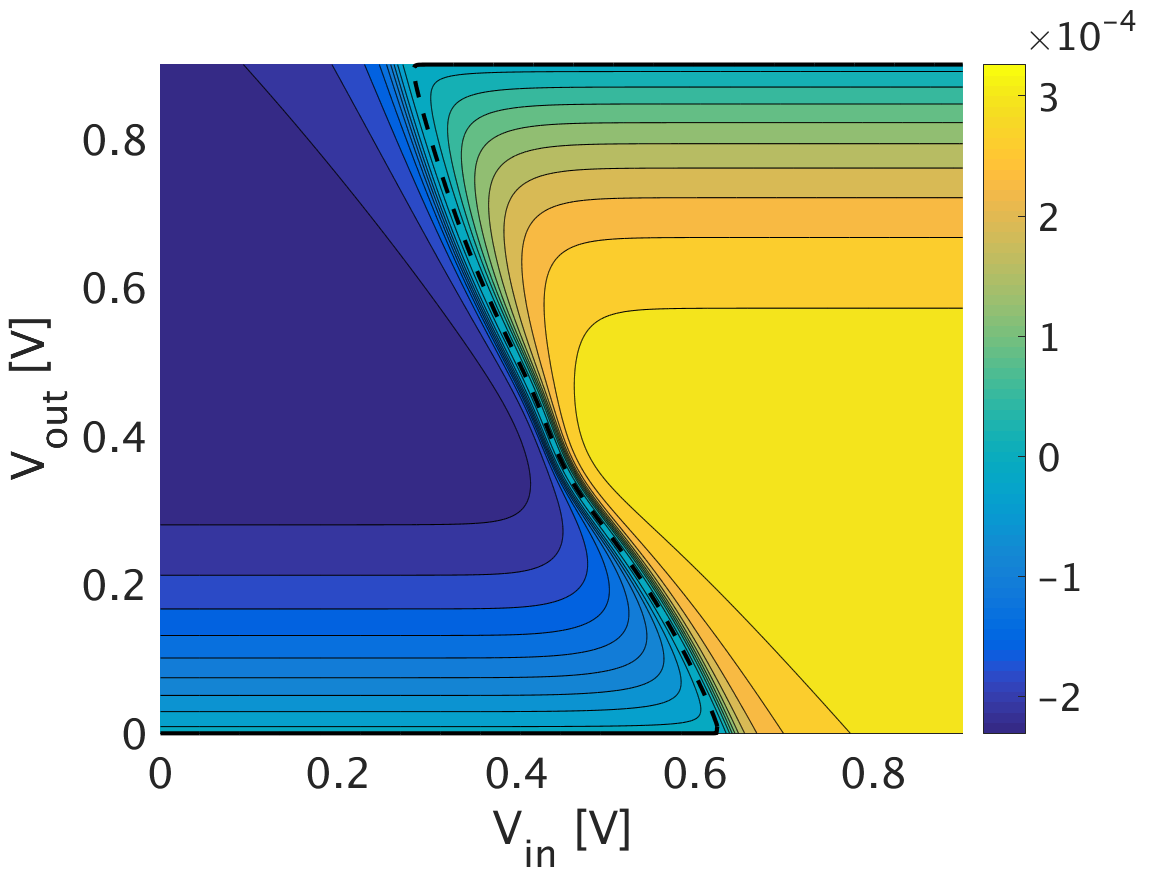}{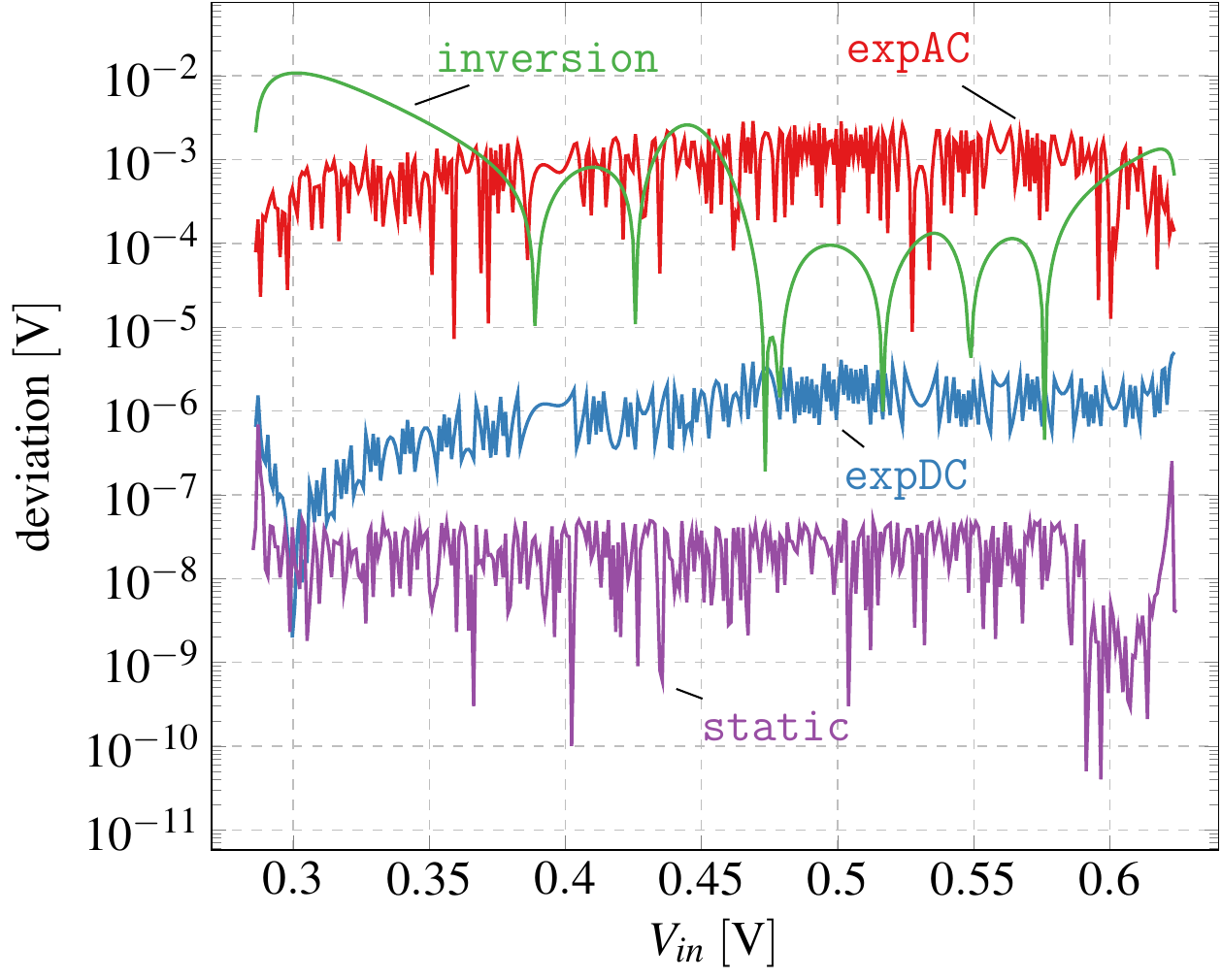}{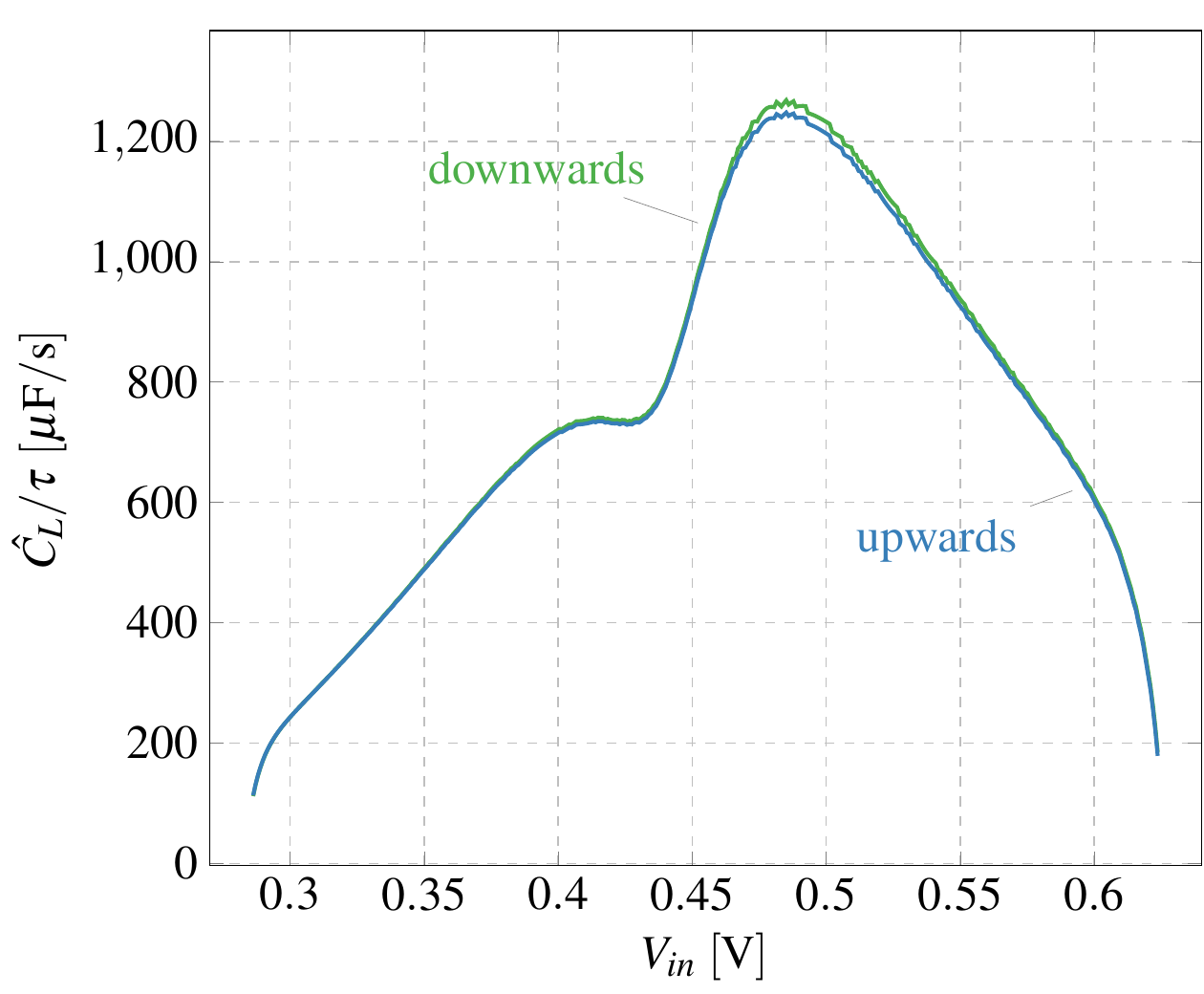}{0.30}

Since the target of $\pm \SI{10}{\uV}$ was quite deliberately chosen, let us review
the impact of target accuracy on the run time for each method:

\begin{description}[align=left]
\item[\binary] The amount of binary steps has hardly any impact on the
  runtime. We experienced a reduction by only $10$ \% when switching from
  $40$ to $20$ iterations with a accuracy loss of four orders of magnitude.
\item[\map] The run time scales linearly with the amount of points which is quite
  natural as each one is determined by a separate \dc\ analysis.
\item[\tranEst\ \& \dcEst] The accuracy increases only with a more accurate
  \map. Apart from that, their simulation time is constant.
\item[\invTran] The chosen gain $p$ of the current source has a high impact
  on the simulation time. Higher gain decreases run time
  but also yields more instabilities, e.g. oscillations. Increasing the simulation
  period helps to resolve these, however the run time increases almost by the
  same factor.
\item[\invDC] The simulation time is constant, and always the
  highest possible accuracy is delivered. In our setting, however, the output data format
  limited the attainable (exportable) accuracy.
\end{description}

\noindent Clearly there is still room for optimizations such as
\begin{itemize}
\item determining $\iout$ solely for grid points close to the metastable line for \map
\item better choice of the gain $p$ in \invTran\ or even improving
  the whole control loop dynamics by an integral or differential part
\item non-uniform simulation time for \invTran
\end{itemize}
which we left for future research.
For this reason the presented run times shall only be used to get an intuition how
long the characterization approximately takes. 

Since we use both transient and \dc\ analyses overall we experienced that the
former are much harder to handle, as more parameters have to be defined,
primarily the time period of simulation. In addition further complications such
as cutting the first part of the simulation in \tranEst\ or an appropriate gain
for the current source in \invTran\ have to be overcome. In total \dc\ analyses
achieve better results in shorter time with simpler methods. During our research
we even realized that most methods, in particular \invDC, are also applicable to
other problems, such as determining the metastable value of a flip-flop. We
consider this more a lucky coincidence than a designed feature as it is a side
effect of the utilized \emph{Newton-Raphson} algorithm.

\begin{figure}[b]
  \centering
  \subfloat[\std \label{fig:standard6T}]{%
    \begin{minipage}[b][4.3cm]{0.445\linewidth}
      \centering
      \includegraphics[width=1\linewidth]{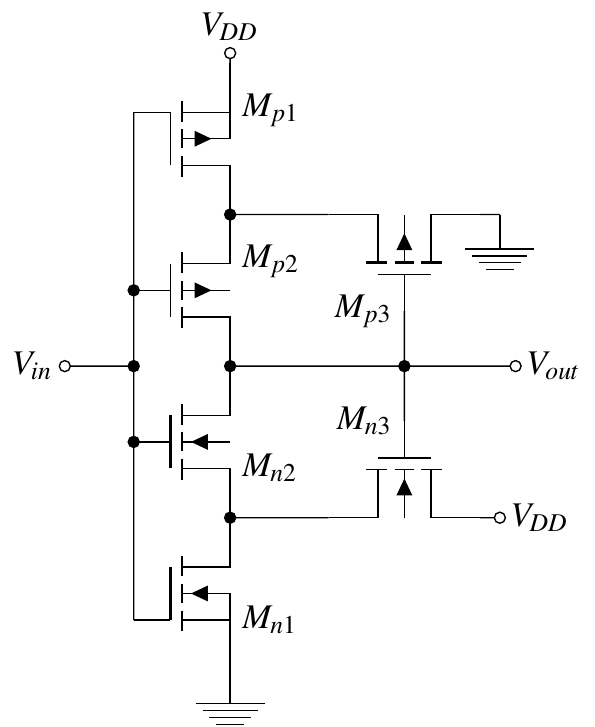}
    \end{minipage}
  }
  \subfloat[\invLoop \label{fig:inverter_loop}]{%
    \begin{minipage}[b][4.3cm]{0.535\linewidth}
      \centering
      \includegraphics[width=1\linewidth]{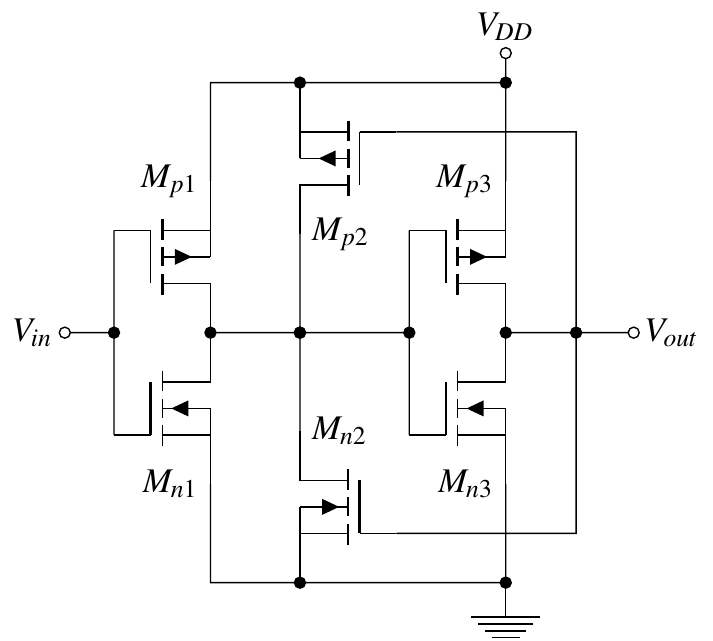}
    \end{minipage}
  }
  \caption{Transistor level circuit implementations.}
\end{figure}

\subsection{Standard Implementation (\std)}
\label{sec:standard6T}

To compare our results with those from Steininger et al. we first analyze the
implementation used in~\cite{SMN16:ASYNC}. The
transistor level circuit is shown in Fig.~\ref{fig:standard6T}.

\compPlot{sim_ctrl_hyst}{\ctrlHyst}{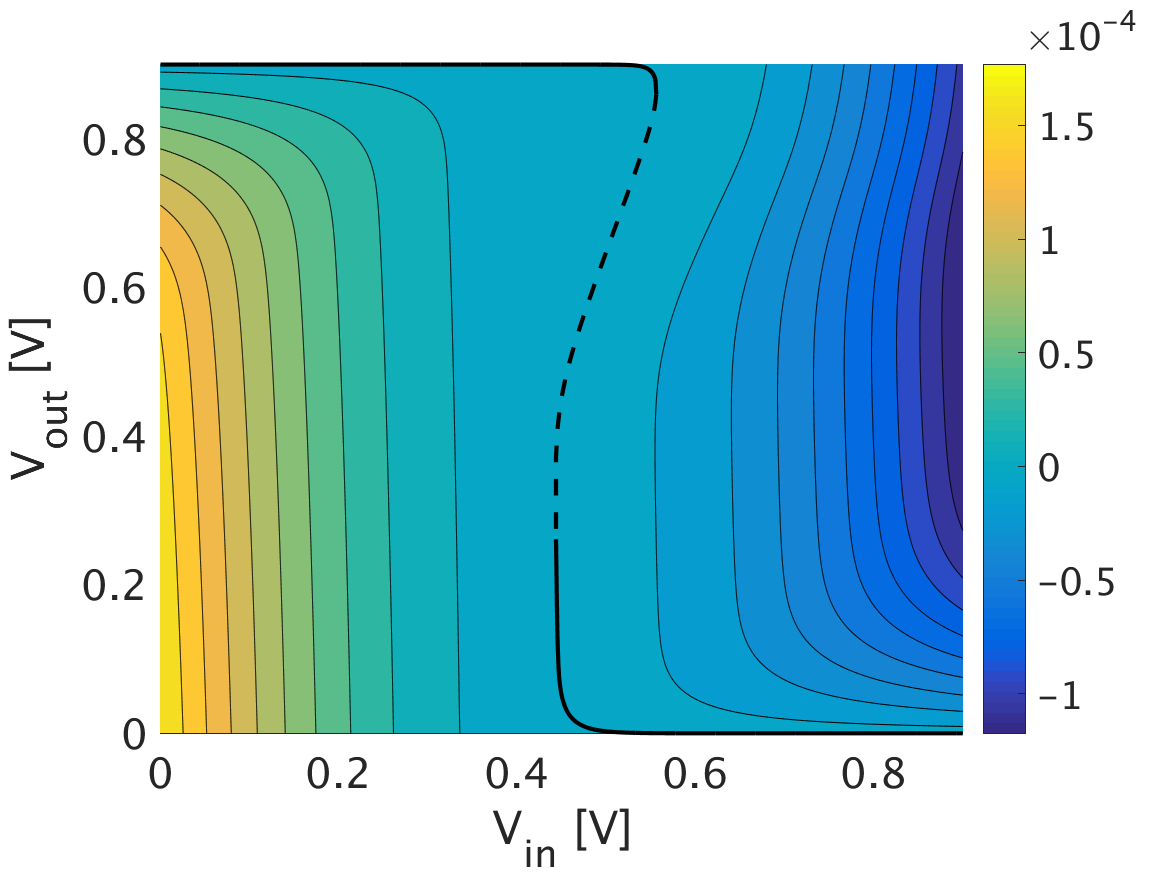}{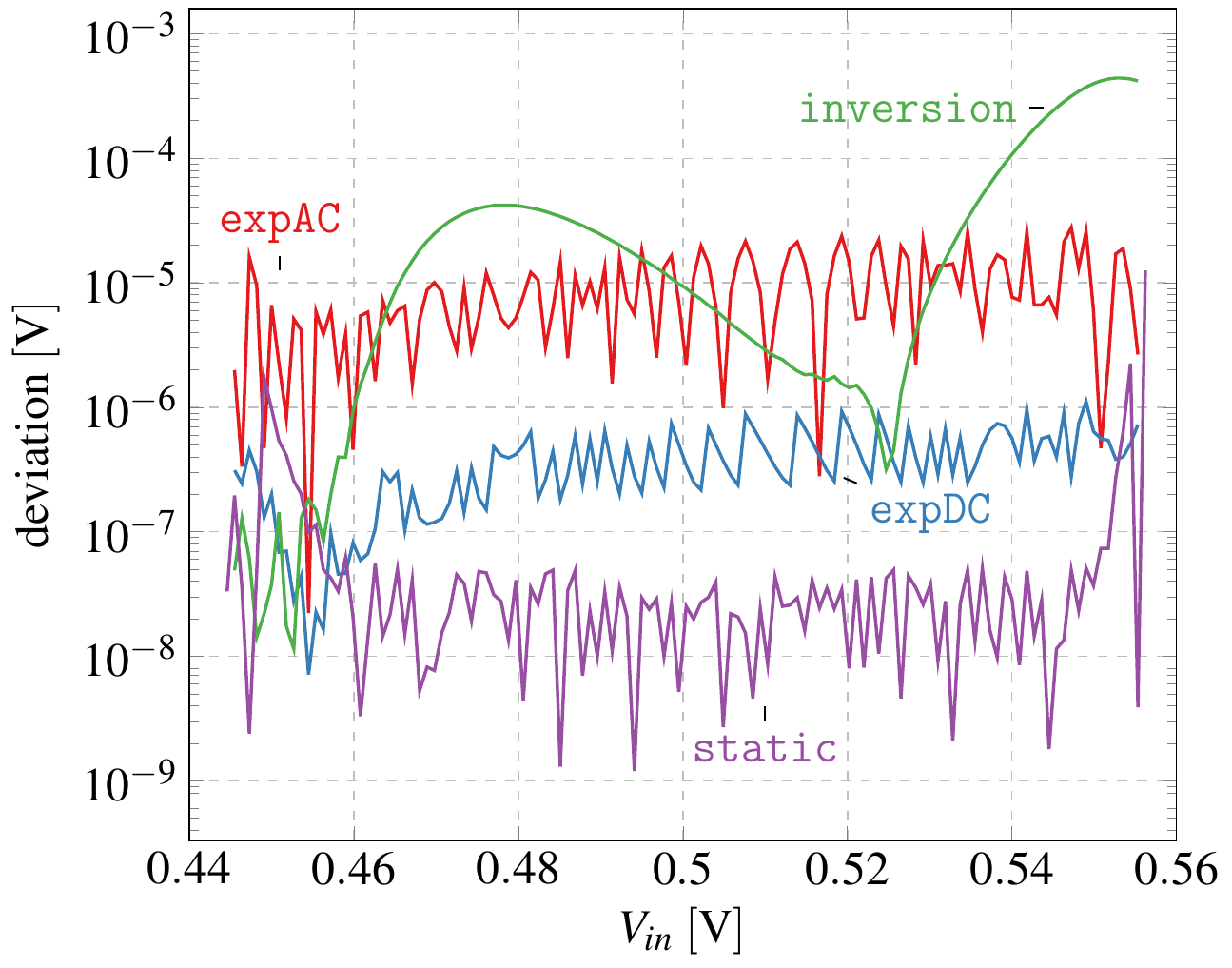}{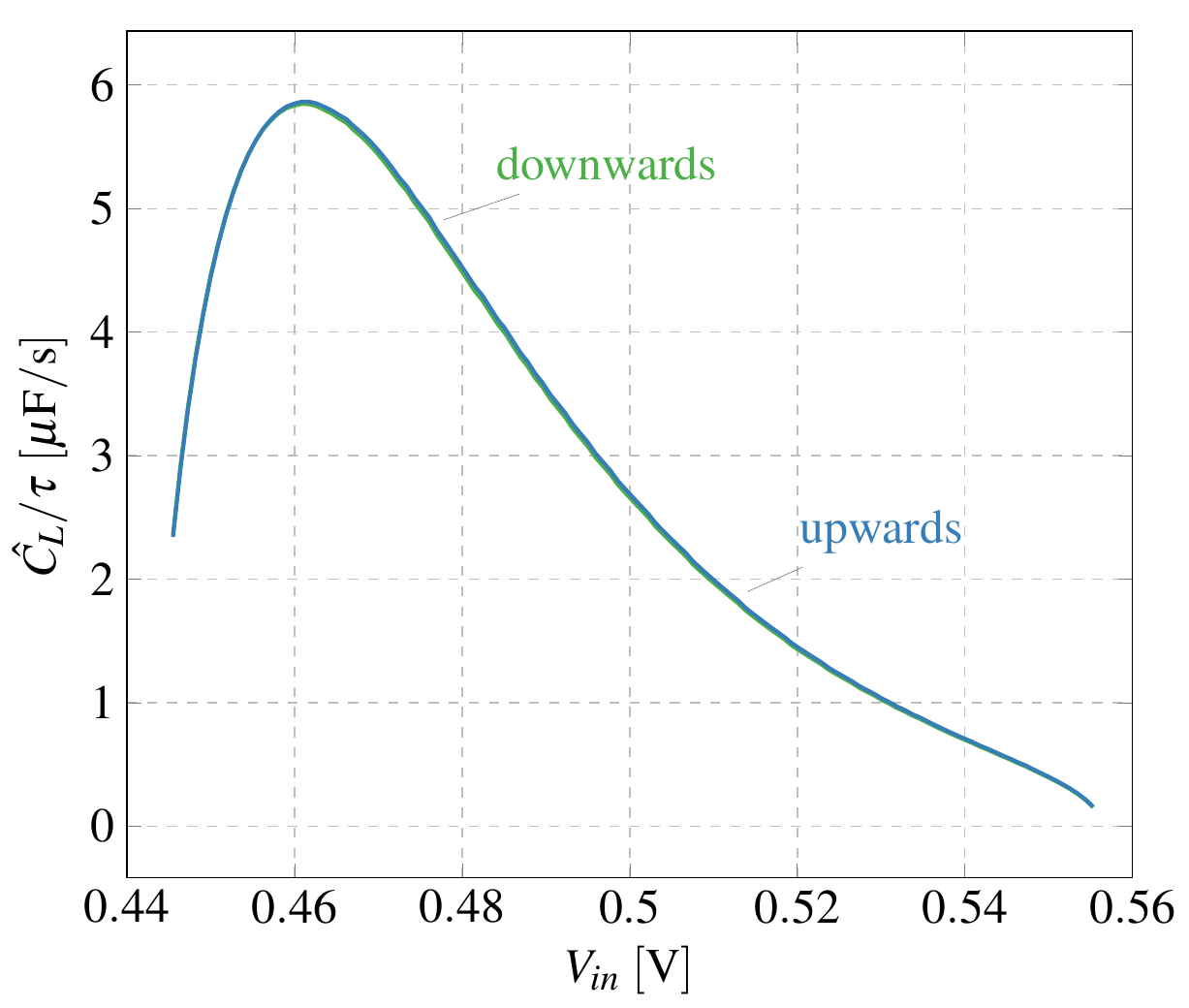}{0.31}

The achieved (meta-)stable line (shown in Fig.~\ref{fig:sim_standard6T_dVout},
$\gone$ and $\gthree$ solid, $\gtwo$ dashed) fits very well to the results
published in~\cite{SMN16:ASYNC}.  Thus we conclude that our tool works as
expected. The same figure also shows a heat map of the output current $\iout$ in
the $\vin$-$\vout$ plane. Please note that the $\iout$-spacing of the contour
lines is linear.  This means that close to the metastable line $\iout$ changes
only moderately, as expected from the exponential resolution trajectories
predicted by theory.

In contrast to the calculations of Marino~\cite{Marino77}, however, whose
$\vout'$ only depends on the distance to the final, stable state, our results
show additional dependencies of $\iout$.  This can be seen very clearly by
observing its maximum and minimum, which both are near $\vdd/2$.

Fig.~\ref{fig:sim_standard6T_dev} shows the (absolute) deviation of the
predicted metastable voltages from the ones of \binary\ .  Please note that with
the finite export number format deviations below \SI{5e-8}{\V} were out of
reach. From our result we therefore deduce that $\invDC$ and $\binary$ are
capable of delivering the same accuracy.

Finally \ref{fig:sim_standard6T_tau} shows the (inverse of the) resolution
constant $\hat{C_L}/\tau$ determined by \dcEst. It significantly varies with
$\vin$, with the biggest (best) value in the middle and the lowest at points
close to the stable states, meaning that the latter are left more slowly. The
values achieved for the up- and down-resolving waveform are shown separately,
but the graphs nicely overlap.

The significant change of $\tau$ raises the interesting question whether the
quick resolution from the middle will cross the far distance to the saturation
faster than the slow one from the borders that only has a short distance to
cross. In consequence one might determine a worst starting point from which
resolution takes the longest. However, the answer heavily depends on what is
considered the threshold for ``resolved'', and, most importantly, on how deep
the initial metastability was (recall that resolution time is essentially
unbounded).

\begin{figure}[h]
  \centering
  \includegraphics[width=0.9\linewidth]{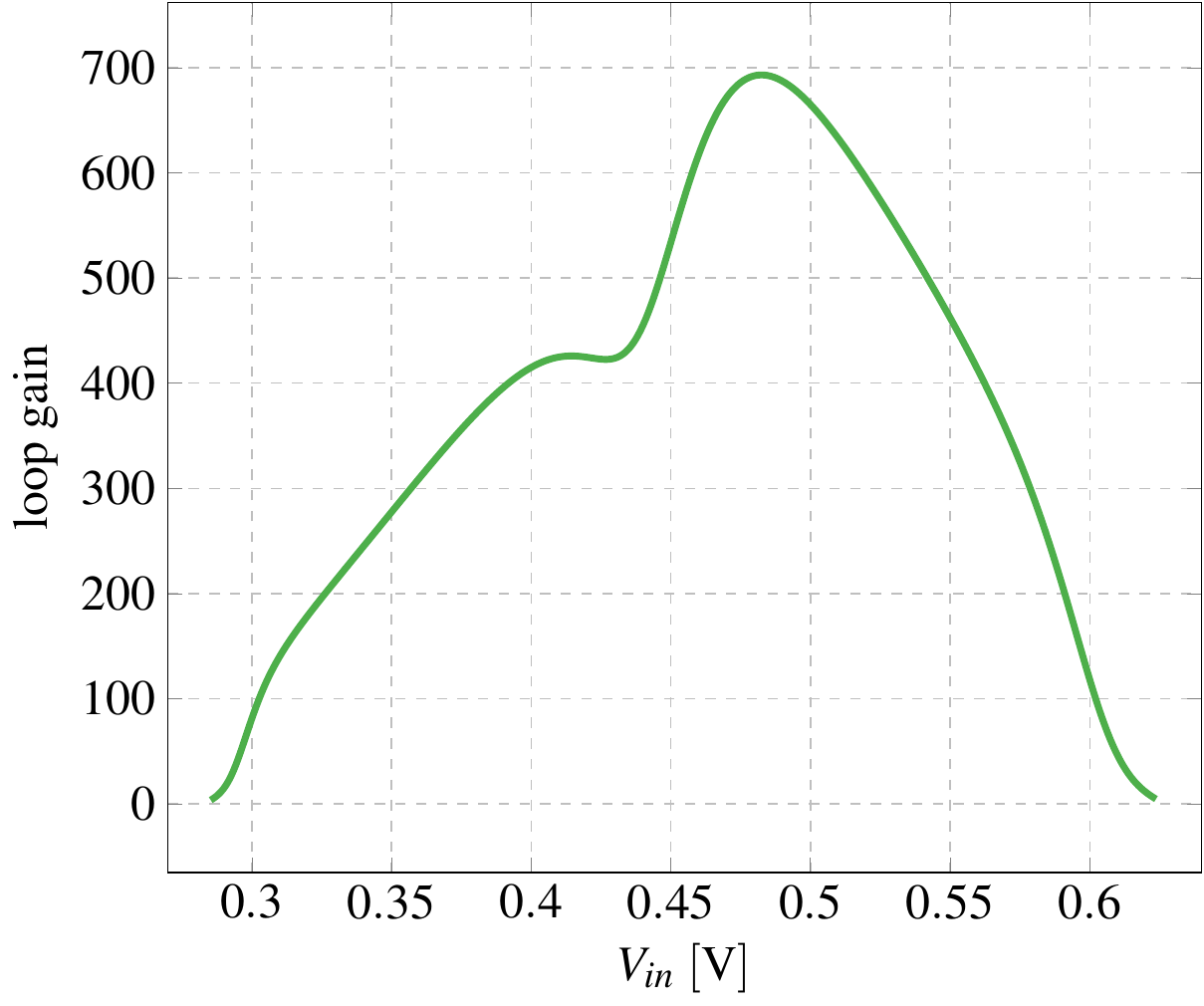}
  \caption{Memory loop gain of \invLoop.}
  \label{fig:inv_loop_amp}
\end{figure}

\subsection{Inverter Loop (\invLoop)}
\label{sec:inverter_loop}

The second circuit we investigate (transistor level implementation see
Fig.~\ref{fig:inverter_loop}) is essentially a latch whose input can not be
decoupled any more, i.e., a plain inverter loop (preceded by an additional
inverter).  The hysteresis is defined by the relation between the driving
strength of the first inverter (transistors $M_{p1}$ and $M_{n1}$) and the weak
feed back one ($M_{p2}$ and $M_{n2}$). For the latter we reduced the width to
one tenth.

The $\iout$ map (see Fig.~\ref{fig:sim_inv_loop_dVout}) significantly differs
from the one of \std. First of all the contour lines are horizontal, much more
like the prediction made by Marino in~\cite{Marino77}.  Secondly the current
changes much more rapidly than before, which also leads to much higher values
for $\hat{C_L}/\tau$ (three orders of magnitude, see
Fig.~\ref{fig:sim_inv_loop_tau}), i.e., metastability is resolved much quicker.
According to \cref{equ:p} $p$ can be calculated using the memory loop gain shown
in \cref{fig:inv_loop_amp}. Picking the highest value ($A^2=700$) results in
$p=1.000357$ which was confirmed to be a suitable value by
simulations. Increasing $p$ however quickly leads to oscillations of $\vout$,
whereat metastable points with higher loop amplification become instable
earlier.

\begin{figure}[b]
  \centering
  \includegraphics[width=0.75\linewidth]{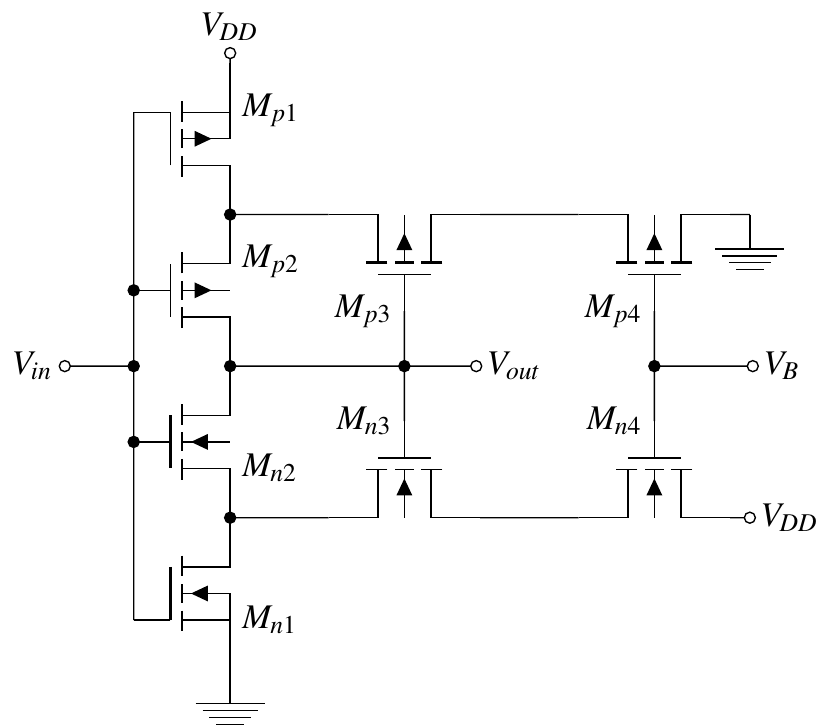}
  \caption{Circuit implementation of \ctrlHyst.}
  \label{fig:adjustable_hysteresis}
\end{figure}

\subsection{Adjustable Hysteresis (\ctrlHyst)}
\label{sec:adjustable_hysteresis}

In some applications it is important to adjust the hysteresis of the \gls{st}
during operation. One circuit that can be used for this purpose is shown in
Fig.~\ref{fig:adjustable_hysteresis}. The value $V_B$ on an additional input
alters the position and width of the hysteresis. In our simulations we used
$V_B=\vdd$ as in this case the hysteresis is the widest and thus the most stable
one. Anyway, similar behavior could be observed for other choices of $V_B$ as
well.

The first remarkable thing in Fig.~\ref{fig:sim_ctrl_hyst_dVout} is the
relatively large peak value of $\vout$ on $\gthree$.  It reaches up to about
\SI{0.3}{\V} which is one third of the supply voltage and almost certainly in
the forbidden region.  Please recall (see Section~\ref{sec:hysteresis}) that
those states can be easily reached by a ramp stopping at a defined value,
implying that resilience against metastability is weakened a lot. The \map\
furthermore reveals nearly vertical contour lines in the left half of the
plot. This suggests, that in this region the output slope is constant, i.e., we
get ramps at the output. Very surprisingly, $\iout$ does not seem to depend on
the distance to the stable state at all, as it was calculated by
Marino~\cite{Marino77}.

The values of $\hat{C_L}/\tau$ (Fig.~\ref{fig:sim_ctrl_hyst_tau}) are comparable
to \std~.  However, since the stable states from below reach far into the
mid-voltage region the graph is not symmetric any more.

\section{Conclusion and Future Work}
\label{sec:conclusion}

In this paper we have presented several ways to characterize the metastable
behavior of a \glsdesc{st} including estimations that increase accuracy beyond
numerical precision, a novel method to convert metastable states into stable
ones, and plain \dc\ analysis that turned out to be already sufficient to
accurately determine metastability. By applying them to three common CMOS
implementations we not only verified that they work properly but were also able
to compare them, giving an edge to \dc\ rather than transient methods regarding
accuracy and run time. At the same time simulation results revealed that the
metastable behavior only partially follows the theoretical predictions made in
the past.

For future work we want to use the results of this paper as starting point to
derive an expression for estimating the reliability impact of metastable upsets
in \glspl{st}, comparable to the MTBU formula in flip-flops.

\input{ms.bbl}

\end{document}

%% file: common.tex
\newcommand{\hspice}{\textit{HSPICE}}
\newcommand{\spice}{\textit{SPICE}}

\newcommand{\dc}{\textit{DC}}
\newcommand{\ac}{\textit{AC}}

\newcommand{\vdd}{V_{DD}}
\newcommand{\gnd}{\textit{GND}}
\newcommand{\vout}{V_{out}}
\newcommand{\vin}{V_{in}}
\newcommand{\vl}{V_{L}}
\newcommand{\vh}{V_{H}}
\newcommand{\vmeta}{V_{M}}

\newcommand{\iout}{I_{out}}

\newcommand{\inmos}{I_{n}}
\newcommand{\ipmos}{I_{p}}

\newcommand{\std}{\texttt{std}}
\newcommand{\invLoop}{\texttt{loop}}
\newcommand{\ctrlHyst}{\texttt{adjust}}

\newcommand{\hyst}{\texttt{hyst}}
\newcommand{\binary}{\texttt{binary}}
\newcommand{\map}{\texttt{map}}
\newcommand{\tranEst}{\texttt{expAC}}
\newcommand{\dcEst}{\texttt{expDC}}
\newcommand{\invTran}{\texttt{inversion}}
\newcommand{\invDC}{\texttt{static}}

\newcommand{\gone}{\gamma_1}
\newcommand{\gtwo}{\gamma_2}
\newcommand{\gthree}{\gamma_3}

\newcommand{\compPlot}[6]{
\FPeval\result{0.6-#6}
\begin{figure*}[t!]
  \centering
  \subfloat[\map \label{fig:#1_dVout}]{%
    \begin{minipage}[t][4.2cm][c]{0.3\linewidth}
      \centering
      \includegraphics[width=\linewidth]{#3}
    \end{minipage}
  }
  \subfloat[$\vmeta$ predictions vs. \binary\label{fig:#1_dev}]{%
    \begin{minipage}[t][4.2cm][c]{#6\linewidth}
      \centering
      \includegraphics[width=\linewidth]{#4}
    \end{minipage}
  }
  \subfloat[resolution constant (\dcEst)\label{fig:#1_tau}]{%
    \begin{minipage}[t][4.2cm][c]{\result\linewidth}
      \centering
      \includegraphics[width=\linewidth]{#5}
    \end{minipage}
  }
  \caption{Simulation results for #2.}
  \label{fig:#1}
\end{figure*}
}